\documentclass{aa}  
\usepackage{graphicx}
\usepackage{hyperref}
\usepackage{txfonts}
\usepackage{physics}
\usepackage{siunitx}
\usepackage{float}
\usepackage[table]{xcolor}
\usepackage{amsmath}
\usepackage{orcidlink}
\newcommand{\kvec}{{\vec{k}}}

\newcommand{\sinc}{\mathrm{sinc}}

\begin{document} 
\setlength{\abovecaptionskip}{3pt}

   \title{COMAP Pathfinder -- Season 2 results\\ I. Improved data selection and processing}
   \authorrunning{J.~G.~S.~Lunde et al.}
   \titlerunning{COMAP Pathfinder -- Season 2 results\\ I. Improved data selection and processing}
   \author{
        J.~G.~S.~Lunde\inst{1}\fnmsep\thanks{\email{\href{mailto:j.g.s.lunde@astro.uio.no}{j.g.s.lunde@astro.uio.no}}}\orcidlink{0000-0002-7091-8779}
        \and 
        N.-O.~Stutzer\inst{1}\orcidlink{0000-0001-5301-1377}
        \and 
        P.~C.~Breysse\inst{6}\fnmsep\inst{7}\orcidlink{0000-0001-8382-5275}
        \and 
        D.~T.~Chung \inst{3}\fnmsep\inst{4}\fnmsep\inst{5}\orcidlink{0000-0003-2618-6504}
        \and 
        K.~A.~Cleary\inst{2}\orcidlink{0000-0002-8214-8265}
        \and 
        D.~A.~Dunne\inst{2}\orcidlink{0000-0002-5223-8315}
        \and 
        H.~K.~Eriksen\inst{1}\orcidlink{0000-0003-2332-5281}
        \and
        S.~E.~Harper\inst{10}\orcidlink{0000-0001-7911-5553}
        \and 
        H.~T.~Ihle\inst{1}\orcidlink{0000-0003-3420-7766}
        \and 
        J.~W.~Lamb\inst{9}\orcidlink{0000-0002-5959-1285}
        \and 
        T.~J.~Pearson\inst{2}\orcidlink{0000-0001-5213-6231}
        \and 
        L.~Philip\inst{12}\fnmsep\inst{16}\orcidlink{0000-0001-7612-2379}
        \and
        I.~K.~Wehus\inst{1}\orcidlink{0000-0003-3821-7275}
        \and
        D.~P.~Woody\inst{9}
        \and
        J.~R.~Bond\inst{3}\fnmsep\inst{14}\fnmsep\inst{15}\orcidlink{0000-0003-2358-9949}
        \and 
        S.~E.~Church\inst{17}
        \and 
        T.~Gaier\inst{12}
        \and 
        J.~O.~Gundersen\inst{18}\orcidlink{0000-0002-7524-4355}
        \and 
        A.~I.~Harris\inst{19}\orcidlink{0000-0001-6159-9174}
        \and 
        R.~Hobbs\inst{9}
        \and 
        J.~Kim\inst{13}\orcidlink{0000-0002-4274-9373}
        \and
        C.~R.~Lawrence\inst{12}\orcidlink{0000-0002-5983-6481}
        \and
        N.~Murray\inst{3}\fnmsep\inst{14}\fnmsep\inst{15}
        \and
        H.~Padmanabhan\inst{8}\orcidlink{0000-0002-8800-5740}
        \and
        A.~C.~S.~Readhead\inst{2}\orcidlink{0000-0001-9152-961X}
        \and 
        T.~J.~Rennie\inst{10}\fnmsep\inst{11}\orcidlink{0000-0002-1667-3897}
        \and 
        D.~Tolgay\inst{3}\fnmsep\inst{14}\orcidlink{0000-0002-3155-946X}
        {(COMAP Collaboration)}
    }

   \institute{
       Institute of Theoretical Astrophysics, University of Oslo, P.O. Box 1029 Blindern, N-0315 Oslo, Norway
       \and 
       California Institute of Technology, 1200 E. California Blvd., Pasadena, CA 91125, USA
       \and 
       Canadian Institute for Theoretical Astrophysics, University of Toronto, 60 St. George Street, Toronto, ON M5S 3H8, Canada
       \and 
       Dunlap Institute for Astronomy and Astrophysics, University of Toronto, 50 St. George Street, Toronto, ON M5S 3H4, Canada
       \and 
       Department of Astronomy, Cornell University, Ithaca, NY 14853, USA
       \and 
       Center for Cosmology and Particle Physics, Department of Physics, New York University, 726 Broadway, New York, NY, 10003, U.S.A 
       \and 
       Department of Physics, Southern Methodist University, Dallas, TX 75275, USA
       \and 
       Departement de Physique Théorique, Universite de Genève, 24 Quai Ernest-Ansermet, CH-1211 Genève 4, Switzerland
       \and 
       Owens Valley Radio Observatory, California Institute of Technology, Big Pine, CA 93513, USA
       \and 
       Jodrell Bank Centre for Astrophysics, Alan Turing Building, Department of Physics and Astronomy, School of Natural Sciences, The University of Manchester, Oxford Road, Manchester, M13 9PL, U.K.
       \and 
        Department of Physics and Astronomy, University of British Columbia, Vancouver, BC Canada V6T 1Z1, Canada
        \and 
        Jet Propulsion Laboratory, California Institute of Technology, 4800 Oak Grove Drive, Pasadena, CA 91109
        \and
        Department of Physics, Korea Advanced Institute of Science and Technology (KAIST), 291 Daehak-ro, Yuseong-gu, Daejeon 34141, Republic of Korea
        \and 
        Department of Physics, University of Toronto, 60 St.~George Street, Toronto, ON, M5S 1A7, Canada
        \and 
        David A.~Dunlap Department of Astronomy, University of Toronto, 50 St.~George Street, Toronto, ON, M5S 3H4, Canada
        \and 
        Brookhaven National Laboratory, Upton, NY 11973-5000
        \and 
        Kavli Institute for Particle Astrophysics and Cosmology and Physics Department, Stanford University, Stanford, CA 94305, USA
        \and 
        Department of Physics, University of Miami, 1320 Campo Sano Avenue, Coral Gables, FL 33146, USA
        \and 
        Department of Astronomy, University of Maryland, College Park, MD 20742
   }

   \date{Received 14 June 2024 / Accepted 1 October 2024}

    \abstract
    {
      The CO Mapping Array Project (COMAP) Pathfinder is performing line intensity mapping of CO emission to trace the distribution of unresolved galaxies at redshift $z \sim 3$. We present an improved version of the COMAP data processing pipeline and apply it to the first two seasons of observations. This analysis improves on the COMAP Early Science (ES) results in several key aspects. On the observational side, all second season scans were made in constant-elevation mode, after noting that the previous Lissajous scans were associated with increased systematic errors; those scans accounted for 50\% of the total Season 1 data volume. In addition, all new observations were restricted to an elevation range of 35--65 degrees to minimize sidelobe ground pickup. On the data processing side, more effective data cleaning in both the time and map domain allowed us to eliminate all data-driven power spectrum-based cuts. This increases the overall data retention and reduces the risk of signal subtraction bias. However, due to the increased sensitivity, two new pointing-correlated systematic errors have emerged, and we introduced a new map-domain PCA filter to suppress these errors. Subtracting only five out of 256 PCA modes, we find that the standard deviation of the cleaned maps decreases by 67\% on large angular scales, and after applying this filter, the maps appear consistent with instrumental noise. Combining all of these improvements, we find that each hour of raw Season 2 observations yields on average 3.2 times more cleaned data compared to the ES analysis. Combining this with the increase in raw observational hours, the effective amount of data available for high-level analysis is a factor of eight higher than in the ES analysis. The resulting maps have reached an uncertainty of $25$--$\SI{50}{\mu K}$ per voxel, providing by far the strongest constraints on cosmological CO line emission published to date. 
    }


   \maketitle

\clearpage
\section{Introduction}
Line intensity mapping (LIM) is an emerging observational technique in which the integrated spectral line emission from many unresolved galaxies is mapped in 3D as a tracer of the cosmological large-scale structure \citep[e.g.,][]{kovetz:2017,kovetz:2019}. It represents a promising and complementary cosmological probe to, say, galaxy surveys and cosmic microwave background (CMB) observations. In particular, LIM offers the potential to survey vast cosmological volumes at high redshift in a manner that is sensitive to emission from the entire galaxy population, not just the brightest objects, as is the case for high-redshift galaxy surveys \citep{Bernal_2022}. The most widely studied emission line for LIM purposes is the 21-cm line of neutral hydrogen \citep{Loeb_2004, Bandura_2014, santos2017meerklass}, which is the most abundant element in the universe, but other high-frequency emission lines also appear promising
\citep{korngut2018spherex, Pullen_2023, akeson2019wide, crites2014time, CCAT_Prime_Collaboration_2022, vieira2020terahertz, Karkare_2022,2020-CONCERTO},
in particular due to their different and complementary physical origin as well as lower levels of astrophysical confusion, Galactic foregrounds, and radio frequency interference.

\begin{table*}
    \setlength{\tabcolsep}{8pt} 
    \renewcommand{\arraystretch}{1.3} 
    \caption{\label{tab:data_volume} Overview of COMAP observation season definitions.}
    \begin{tabular}{l l l l}
    \hline
    \hline
       Name      & Dates             & Observing hours & Notes\\
       \hline
       Season 1  & 05/2019 - 08/2020 & 5,200  & Data published in Early Science. Contains 50\% Lissajous, 50\% CES. \\
       Season 2a & 11/2020 - 04/2022 & 7,900  & 100\% CES from this point forward.\\
       Season 2b & 05/2022 - 11/2023 & 4,400  & After azimuth drive slowdown and sampler frequency change.\\
       \hline
    \end{tabular}
    \vspace{2mm}\tablefoot{Season 2 was split into two sub-seasons, respectively denoted 2a and 2b, as the telescope scanning speed was significantly reduced in May 2022 for mechanical reasons.}
\end{table*}

The CO Mapping Array Project (COMAP) represents one example of such an alternative approach and uses CO as the tracer species \citep[see, e.g.,][]{Lidz2011,Pullen2014,Breysse2014}. The COMAP Pathfinder instrument consists of a 19-feed\footnote{We refer to a full detector chain as a "feed."} focal plane array observing at $26$--$\SI{34}{GHz}$ \citep{Lamb_2022} deployed on a \SI{10.4}{m} Cassegrain telescope. This frequency range corresponds to a redshift of $z\sim 2.4$--$3.4$ for the CO(1--0) line, a period during the Epoch of Galaxy Assembly \citep{Li_2016}. The Pathfinder instrument started observing in 2019, and COMAP has previously published results from the first year of data, named Season 1, in the COMAP Early Science (ES) publications \citep{Cleary_2022,Lamb_2022,Foss_2022,Ihle_2022,comap-V,Rennie_2022,Breysse_2022,dunne2024comap}. These ES results provided the tightest constraints on the CO power spectrum in the clustering regime published to date. Since the release of the ES results, the COMAP Pathfinder instrument has continued to observe while also implementing many important lessons learned from Season 1, both in terms of observing strategy and data processing methodology. Combining the observations from both Seasons 1 and 2 and improving the data analysis procedure, the new results improve upon the ES analysis by almost an order of magnitude in terms of power spectrum sensitivity.

This paper is the first of the COMAP Season 2 paper series, and here we present the low-level data analysis pipeline and map-level results derived from the full COMAP dataset as of the end of Season 2 (November 2023). This work builds on the corresponding Season 1 effort as summarized by \cite{Ihle_2022}. The Season 2 power spectrum and null-test results are presented by \citet{Stutzer_2024}, while \citet{chung:2024} discuss their cosmological implications in terms of structure formation constraints. In parallel with the Season 2 CO observations, the COMAP Pathfinder continues to survey the Galactic plane, with the latest results focusing on the Lambda Orionis region \citep{harper2024comap}.

The remainder of this paper is structured as follows: In Sect.~\ref{sec:data} we summarize the changes made to the observational strategy in Season 2 and provide an overview of the current status of data collection and accumulated volume. In Sect.~\ref{sec:pipeline} we summarize our time-ordered data (TOD) pipeline with a focus on the changes since the ES analysis. In Sect.~\ref{sec:mapmaking} we study the statistical properties of the spectral maps produced by this pipeline while paying particular attention to our new map-domain principal component analysis (PCA) filtering and the systematic errors that this filter is designed to mitigate. In Sect.~\ref{sec:data_selection} we present the current data selection methodology and discuss the resulting improvements in terms of data retention in the time, map, and power spectrum domains. In Sect.~\ref{sec:TF} we present updated transfer function estimates and discuss their generality with respect to non-linear filtering. Finally, we summarize and conclude in Sect.~\ref{sec:conclusions}.

\section{Data collection and observing strategy}
\label{sec:data}
This section first briefly summarizes the COMAP instrument and low-level data collection, which is extensively explored in \cite{Lamb_2022}, before exploring the changes made to the telescope and observing strategy between Seasons 1 and 2.

\subsection{Instrument overview}
The COMAP Pathfinder consists of a 19-feed 26--34~GHz spectrometer focal plane array fielded on a \SI{10.4}{m} Cassegrain telescope located at the Owens Valley Radio Observatory (OVRO). At the central observing frequency of 30~GHz, the telescope has a beam full width at half maximum (FWHM) of $4.5'$. The 8~GHz-wide RF signal is shifted to 2--10~GHz, and then split into two 4~GHz bands. The signals from each band are passed on to a separate ``ROACH-2'' field-programmable gate array spectrometer, which further separates the 4~GHz-wide bands into 2~GHz-wide sidebands. The spectrometer outputs 1024 frequency channels for each of the four sidebands, for a spectral resolution of $\sim$2 MHz.

The telescope is also equipped with a vane of microwave-absorbing material, which is temporarily moved into the field of view of the entire feed array before and after each hour-long ``observation'' to provide absolute calibration of the observed signal in temperature units. Each observation is stored in a single HDF5 file containing both the spectrometer output and various housekeeping data, and these files are referred to as ``Level 1'' data. Each observation of one of the three target fields is divided into ten to fifteen numbered ``scans,'' during which the telescope oscillates in azimuth at constant elevation, repointing ahead of the field to start a new scan each time the field has drifted through the scanning pattern.

\subsection{Status of observations}
Table~\ref{tab:data_volume} shows the raw on-sky integration time per season. COMAP Season 1 included 5,200 on-sky observation hours collected from May 2019 to August 2020, while the second season included 12,300 hours collected between November 2020 and November 2023. In these publications, we present results based on a total on-sky integration time of 17,500 hours, a 3.4-fold increase compared to the ES publications.

Several changes were made to the data collection and observing strategy before and during Season 2. Most of these changes came as direct responses to important lessons learned during the Season 1 data analysis and the aim was to increase the net mapping speed, although one was necessary due to mechanical telescope issues during Season 2. Overall, these changes were highly successful, and Season 2 has a much higher data retention than Season 1, which we discuss in Sect. \ref{sec:data_selection}. The most important changes in the Season 2 observing strategy are the following:
\begin{enumerate}
    \item Observations were restricted to an elevation range of $35^\circ$--$65^\circ$ in order to reduce the impact of ground pick-up via the telescope's sidelobe response. As discussed by \citet{Ihle_2022}, the gradient of the ground pickup changes quickly at both lower and higher elevations, and the corresponding observations were therefore discarded in the Season 1 analysis; in Season 2 we avoid these problematic elevations altogether.
    \item Similarly, Lissajous scans were abandoned in favor of solely using constant elevation scans (CES), since \citet{Foss_2022} found elevation-dependent systematic errors associated with the former.
    \item The two frequency detector sub-bands, that previously covered disjoint ranges of 26--30 GHz and 30--34 GHz \citep{Lamb_2022}, were widened slightly, such that they now overlap; this mitigates data loss due to aliasing near the band edges.
    \item The acceleration of the azimuth drive was halved to increase the longevity of the drive mechanism, which started to show evidence of mechanical wear.
\end{enumerate}
The latter two changes were only implemented in the second half of Season 2, and they mark the beginning of what we refer to as Season 2b. These changes are now discussed in greater detail.

\subsection{Restricting the elevation range}\label{sec:elevation}
Sidelobe pickup of the ground is a potentially worrisome systematic error for COMAP, especially since it is likely to be pointing-correlated. Even though ground pickup is primarily correlated with pointing in alt-azimuthal coordinates, the daily repeating pointing pattern of COMAP means there will still be a strong correlation in equatorial coordinates. Analysis of Season 1 observations \citep{Foss_2022}, that ranged from ${\sim}30^\circ$ to ${\sim}75^\circ$ elevation, showed pointing-correlated systematic errors at the highest and lowest elevations.

\begin{figure}
    \centering
    \includegraphics[width=\linewidth]{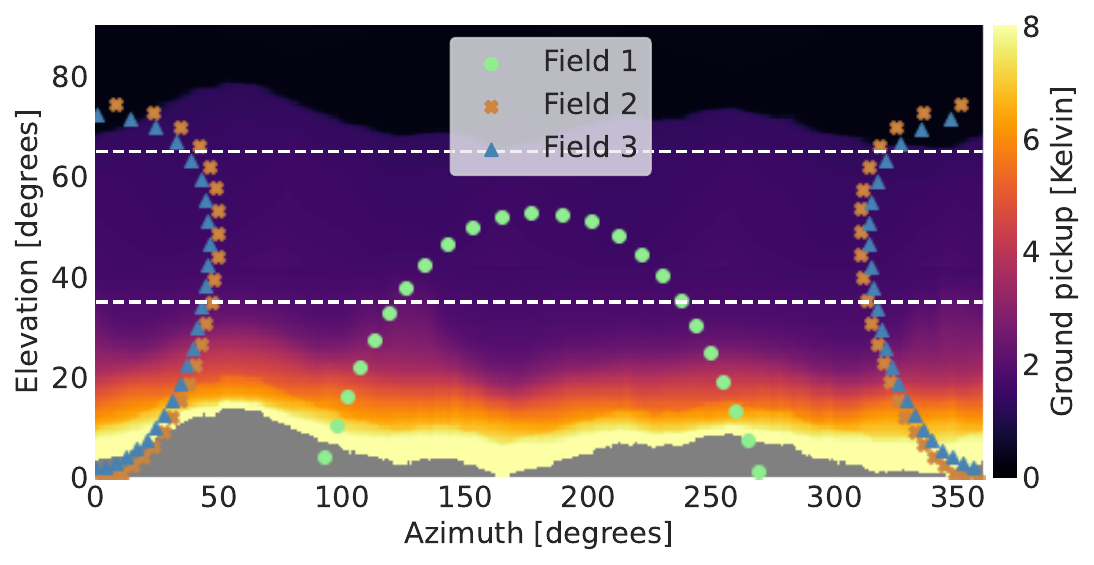}
    \vspace{-4mm}\caption{Approximate sidelobe ground pickup as a function of az and el pointing, simulated by convolving a beam (simulated using GRASP) with the horizon profile (shown in gray) at the telescope site. The paths of the three fields across the sky, in half-hour intervals, are shown, as well as the Season 2 elevation limits at $35^\circ$--$65^\circ$. These limits ensure minimal ground pickup gradient across the field paths.}
    \label{fig:ground_pickup}
\end{figure}

To study this effect in greater detail, we developed a set of antenna beam pattern simulations using GRASP\footnote{https://www.ticra.com/software/grasp/} for the COMAP telescope \citep{Lamb_2022}, and these showed the presence of four sidelobes resulting from the four secondary support legs (SSL), with each sidelobe offset by ${\sim}65^\circ$ from the pointing center. These simulations were convolved with the horizon elevation profile at the telescope site, and the results from these calculations are shown in Fig.~\ref{fig:ground_pickup}. This figure clearly shows that Fields 2 and 3 experience a sharp change in ground pickup around $65^\circ$--$70^\circ$ elevation, as one SSL sidelobe transitions between ground and sky. At very low elevations the ground contribution also varies rapidly for all fields as two of the other SSL sidelobes approach the horizon. While the low-level TOD pipeline removes the absolute signal offset per scan, gradients in the sidelobe pickup over the duration of a scan still lead to signal contamination. We have therefore restricted our observations to the elevation range of $35^\circ$--$65^\circ$, where one SSL sidelobe remains pointed at the ground, and the other three SSL sidelobes are safely pointing at the sky, leaving us with a nearly constant ground pickup. This change incurred little loss in observational efficiency, as almost all allocated observational time outside the new range could be reallocated to other fields within the range.

\subsection{Abandoning Lissajous scans}\label{sec:Lissajous}
The first season of observations contained an even distribution of Lissajous and constant elevation scans (CES), with the aim of exploring the strengths and weaknesses of each. The main strength of the Lissajous scanning technique is that it provides excellent cross-linking by observing each pixel from many angles, which is useful for suppressing correlated noise with a destriper or maximum likelihood mapmaker. The main drawback of this observing mode is that the telescope elevation varies during a single scan, resulting in varying atmosphere and ground pick-up contributions. In contrast, the telescope elevation remains fixed during a CES, producing a simpler pick-up contribution although with somewhat worse cross-linking properties.

When analyzing the Season 1 power spectra resulting from each of the two observing modes, \cite{Ihle_2022} find that the Lissajous data both produce a highly significant power spectrum, especially on larger scales and fail key null tests. The CES scans, on the other hand, produce a power spectrum consistent with zero, and pass the same null tests. We therefore conclude that the significance in the Lissajous power spectrum is due to residual systematic errors. Additionally, the main advantage of Lissajous scanning, namely better cross-linking, prove virtually irrelevant because of a particular feature of the COMAP instrument: because all frequency channels in a single COMAP sideband are processed through the same backend, the instrumental $1/f$ gain fluctuations are extremely tightly correlated across each sideband. As a result, the low-level TOD pipeline is capable of simultaneously removing virtually all correlated noise from both gain and atmosphere by common-mode subtraction (see Sect.~\ref{sec:l2gen:polyfilter}). At our current sensitivity levels, we therefore find no need to employ a complex mapmaking algorithm that fully exploits cross-linking observations, but we can rather use a computationally faster binned mapmaker \citep{Foss_2022}. After Season 1 we therefore concluded that there was no strong motivation to continue with Lissajous scans, and in Season 2 we employ solely CES. 

\subsection{Widening of frequency bands for aliasing mitigation}
\label{sec:aliasing}

As discussed in detail by \citet{Lamb_2022}, the COMAP instrument exhibits a small but non-negligible level of signal aliasing near the edge of each sideband. In the Season 1 analysis, this was accounted for simply by excluding the channels with aliasing power from other channels suppressed by less than $\SI{15}{dB}$. In total, 8\,\% of the total COMAP bandwidth is lost due to this effect, and this leads to gaps in the middle of the COMAP frequency range. Both the origin of the problem and its ultimate solution were known before the Season 1 observations started \citep{Lamb_2022}, but this took time to implement. 

Band-pass filters applied after the first downconversion and low-pass filters applied after the second downconversion allow significant power above the Nyquist frequency into the sampler. This is then aliased into the 0--2.0 GHz observing baseband, requiring the contaminated channels to be excised. By increasing the sampling frequency from 4.000 GHz to 4.250 GHz, the Nyquist frequency is raised to 2.125 GHz, closer to the filter edges. Not only is the amount of aliased power reduced, it is also folded into frequencies above the nominal width of each 2 GHz observing band. The existing samplers were able to accommodate the higher clock speed, but the field-programmable gate array (FPGA) code had to be carefully tuned to reliably process the data. This was finally implemented from the start of Season 2b, and the aliased power is now shifted outside the nominal range of each band, such that the affected channels can be discarded without any loss in frequency coverage. The number of channels across the total frequency range is still 4096, so the ``native'' Season 2b channels are 2.075\,MHz wide, up from 1.953\,MHz.

\subsection{Azimuth drive slowdown}
\label{sec:pointing_speed}

It became clear during Season 2 that the performance of the telescope's azimuth drives had degraded, owing to wear and tear on the drive mechanisms caused by the telescope's high accelerations. In order to protect the drives from damage, the analog acceleration limit was reduced until the stress was judged by its audible signature to be acceptable. Though not carefully quantified, this was about an order of magnitude change, and the minimum time for a scan is therefore about a factor of three less. Additionally, the maximum velocity was reduced by a factor of two in the drive software.

\begin{figure}
    \centering
    \includegraphics[width=\linewidth]{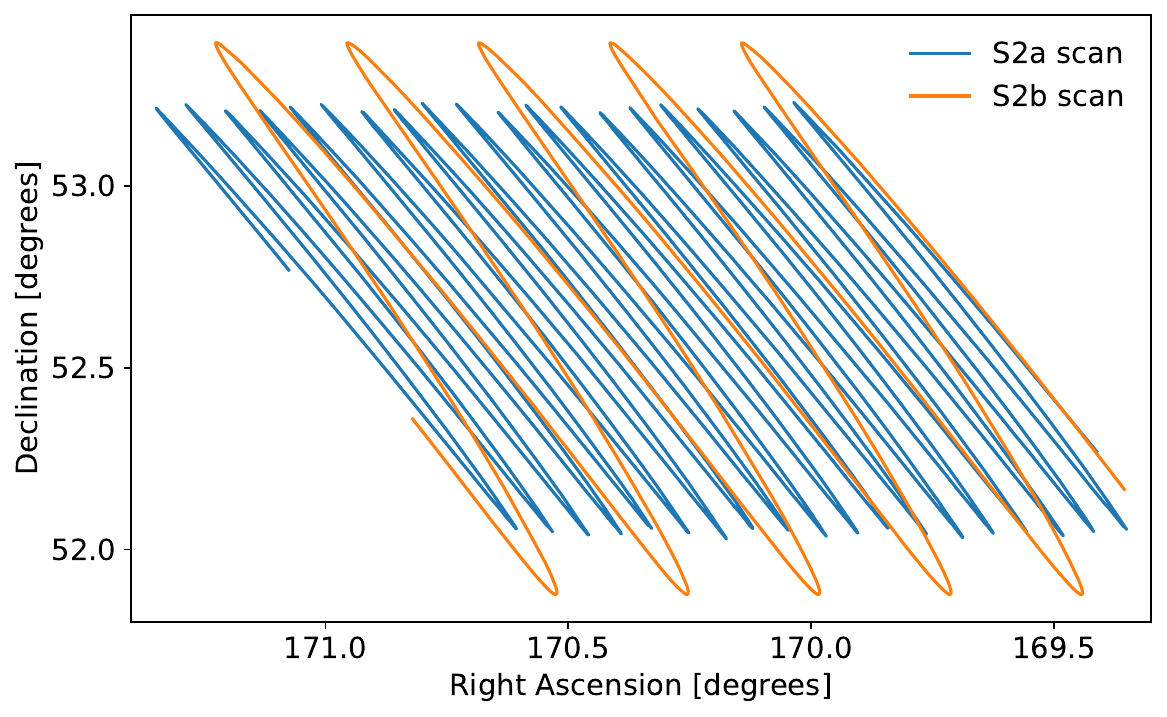}
    \vspace{-4mm}\caption{Comparison of the faster pointing pattern from a Season 2a scan, and the slower pointing pattern of a Season 2b scan. Both patterns show a 5.5-minute constant elevation scan, as the field drifts across.}
    \label{fig:fast_vs_slow_pointing}
\end{figure}

Figure~\ref{fig:fast_vs_slow_pointing} illustrates the old (Season 2a) and new (Season 2b) pointing patterns, with the new pattern being slightly wider and around four times slower. The new realized pointing pattern is now also closer to sinusoidal since the drives are better able to 'keep up' with the sinusoidal pattern of the commanded position, due to the slower velocity.

With the new actually sinusoidal scanning pattern, the integration time is less uniform across each field in each observation, as the telescope spends more time pointing near the edges of the field than it previously did. However, co-adding across the observing season does smooth out the uneven integration time, based on the receiver field of view (each of the 19 feeds observes the sky at a position that is offset from the others) and field rotation (the telescope observes the fields at different angles as they move across the sky).

\subsection{Data storage}
With 19 feeds, 4096 native frequency channels, and a sampling rate of about 50\,samples/sec, COMAP collects 56\,GB/hour of raw 24-bit integer data, stored losslessly as 32-bit floats. The full set of these raw data (combined with telescope housekeeping), named ``Level\,1,'' currently span about 800\,TB. These data are then filtered by our TOD pipeline into the so-called Level\,2 data \citep{Foss_2022}, in which a key step is to co-add the native frequency channels to 31.125 MHz width. These downsampled channels form the basis of the higher-level map-making and power spectrum algorithms. The total amount of Level\,2 data is about 50\,TB. Both Level\,1 and Level\,2 files are now losslessly compressed using the GZIP algorithm \citep{gailly1992gnu}, that achieves average compression factors of 2.4 and 1.4, respectively, reducing the effective sizes of the two datasets to 330 TB and 35 TB. The lower compression factor of the filtered data is expected because the filtering leaves the data much closer to white noise, and therefore with a much higher entropy.

The files are stored in the HDF5 format \citep{koranne2011hierarchical}, which allows seamless integration of compression. Both compression and decompression happen automatically when writing to and reading from each file. GZIP is also a relatively fast compression algorithm, taking roughly one hour to compress each hour of COMAP data on a typical single CPU core. Decompression takes a few minutes per hour of data, which is an insignificant proportion of the total pipeline runtime. HDF5 also allows for arbitrary chunking of datasets before compression. Chunking aids in optimizing performance since the Level~1 files consist of entire observations (1~hour), but the current TOD pipeline implementation (see Sect.~\ref{sec:l2gen_perf}) reads only individual scans of 5~minutes each. We partition the data into chunks of 1-minute intervals to minimize redundant decompression when accessing single scans. Other compression algorithms have been tested, and some, such as lzma\footnote{https://tukaani.org/xz/}, achieve up to a 20\,\% higher compression factor on our data. They are, however, also much slower at both compression and decompression, and they interface less easily with HDF5.

\section{The COMAP TOD pipeline}
\label{sec:pipeline}

\begin{figure*}[th!]
    \centering
    \includegraphics[width=\textwidth]{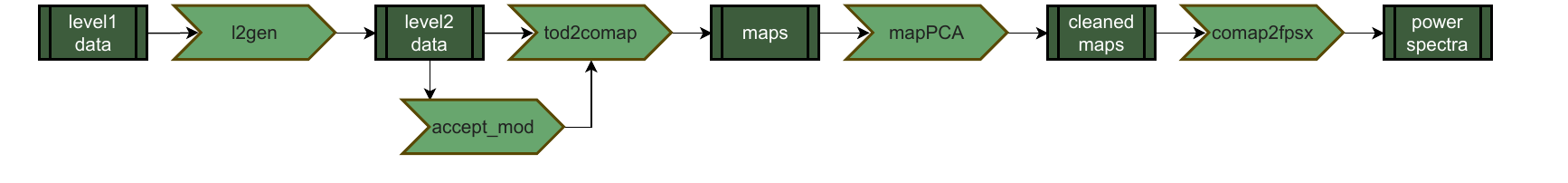}
    \vspace{-4mm}\caption{Flowchart of the COMAP pipeline, from raw Level1 data to final power spectra. Data products are shown as darker boxes and pipeline code as lighter arrows. \texttt{l2gen} performs the time-domain filtering, turning raw data into cleaned Level 2 files. \texttt{accept\_mod} performs scan-level data selection on cleaned data. \texttt{tod2comap} is a simple binned mapmaker. \texttt{mapPCA} performs a map-level PCA filtering. Finally, \texttt{comap2fpsx} calculates power spectra as described by \citet{Stutzer_2024}.}
    \label{fig:flowchart}
\end{figure*}

This section lays out the COMAP time-domain pipeline, named \texttt{l2gen}, focusing on the changes from the first generation pipeline, which is described in detail by \cite{Foss_2022}. The pipeline has been entirely re-implemented (see Sect.~\ref{sec:l2gen_perf}) for performance and maintainability reasons but remains mathematically similar. Figure \ref{fig:flowchart} shows a flowchart of the entire COMAP pipeline, of which \texttt{l2gen} is the first step.

The purpose of the TOD pipeline is to convert raw detector readout (Level 1 files) into calibrated time-domain data (Level 2 files) while performing two key operations: substantially reducing correlated noise and systematic errors, and calibrating to brightness temperature units. COMAP uses a filter-and-bin pipeline, meaning that we perform as much data cleaning as possible in the time domain, before binning the data into maps with na\"{i}ve noise-weighting. This leaves us with a biased pseudo-power spectrum, that can be corrected by estimating the pipeline transfer function (Sect.~\ref{sec:TF}).

The following sections explain the main filters in the TOD pipeline. The normalization (Sec. \ref{sec:l2gen:norm}), $1/f$ filter (Sec. \ref{sec:l2gen:polyfilter}), calibration and downsampling (Sect.~\ref{sec:l2gen:calib}) steps remain unchanged from the ES pipeline, and are briefly summarized for completeness. We denote the data at different stages of the pipeline as $d^\mathrm{\,name}_{\nu,t}$, where the $\nu,t$ subscript indicates data with both frequency and time dependence.

\subsection{System temperature calculation}\label{sec:l2gen:tsys_calc}
The first step of the pipeline is to calculate the system temperature $T^\mathrm{sys}_\nu$ of each channel in the TOD. At the beginning and end of each observation, a calibration vane of known temperature is moved into the field of view of all feeds. The measured power from this ``hot load'', $P^\mathrm{hot}_\nu$, and the temperature of the vane, $T^\mathrm{hot}$, are interpolated between calibrations to the center of each scan. Power from a ``cold load'', $P^\mathrm{cold}_\nu$, is calculated as the average power of individual sky scans. The system temperature is then calculated as (see \cite{Foss_2022} for details)
\begin{equation}\label{eqn:tsys}
    T^\mathrm{sys}_\nu = \frac{T^\mathrm{hot} - T^\mathrm{CMB}}{P^\mathrm{hot}_\nu/P^\mathrm{cold}_\nu - 1}
\end{equation}
under the approximation that the ground, sky, and telescope share the same temperature.

\subsection{Pre-pipeline masking}
In ES, \texttt{l2gen} performed all frequency channel masking toward the end of the pipeline. While some masks are data-driven (specifically driven by the filtered data), others are not. We now apply the latter category of masks prior to the filters, to improve the filtering effectiveness. These are
\begin{itemize}
    \item masking of channels that have consistently been found to be correlated with systematic errors, and have been manually flagged to always be masked;
    \item for data gathered before May 2022, masking of channels with significant aliased power, as outlined in Sect.~\ref{sec:aliasing};
    \item masking of channels with system temperature spikes, as outlined below.
\end{itemize}

\begin{figure}
    \centering
    \includegraphics[width=\linewidth]{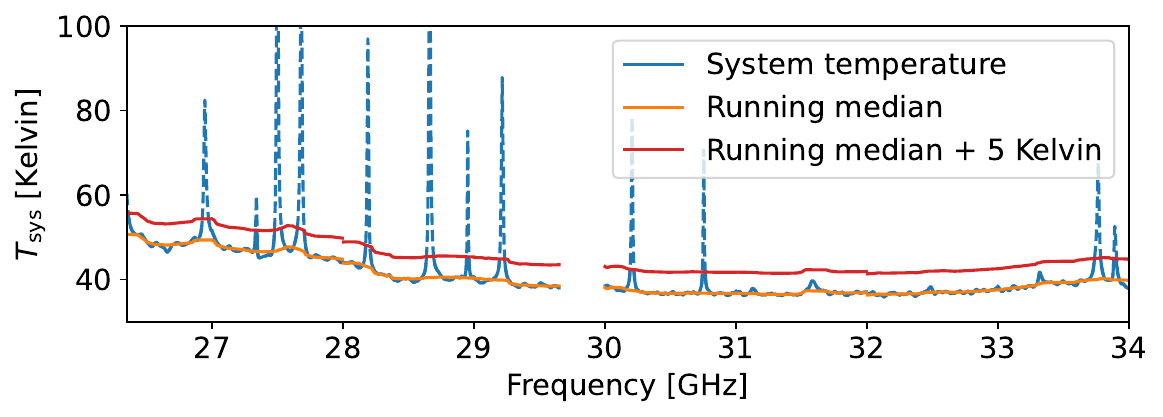}
    \vspace{-4mm}\caption{Example of $T^\mathrm{sys}$ spike masking by running median. The system temperature is shown in blue, the running median in orange, and the 5 Kelvin cut above the running median in red. The $T^\mathrm{sys}$ values which are cut are shown as a dashed instead of solid line.}
    \label{fig:tsys_median_cut}
\end{figure}

The system noise temperature, $T^\mathrm{sys}_\nu$, for each feed's receiver chain has a series of spikes at specific frequencies, believed to result from an interaction between the polarizers and the corrugated feed horn \citep{Lamb_2022}. The spikes are known to be associated with certain systematic errors, and the affected frequency channels are therefore masked out. The ES analysis used a static system temperature threshold of $\SI{85}{K}$ for masking, but the new version of \texttt{l2gen} applies a 400 channel-wide running median kernel to the data and masks all frequencies with a noise temperature of more than 5\,K above the median. We repeat the running median fit and threshold operation once on the masked data, to reduce the impact of the spikes on the fit. The second iteration uses a kernel width of 150 channels. The final running median and threshold are illustrated in Fig.~\ref{fig:tsys_median_cut}. As the system temperature can vary quite a lot across the $\SI{4}{GHz}$ range, this method fits the spikes themselves more tightly, while avoiding cutting away regions of elevated but spike-free system temperature.

The spike frequencies vary from feed to feed, and we are therefore not left with gaps in the redshift coverage of the final 3D maps. On average, we mask $6\%$ of all frequency channels this way. However, because the affected channels are, by definition, more noisy, this only results in a loss of $3\%$ of the sensitivity.

\begin{figure*}
    \centering
    \includegraphics[width=\textwidth]{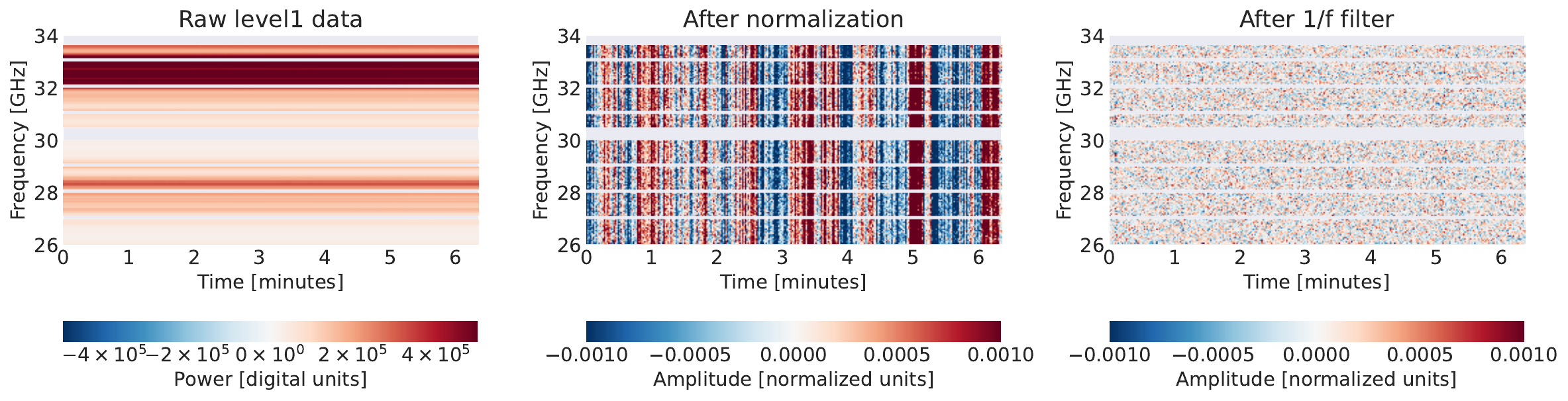}
    \vspace{-4mm}\caption{Illustration of the effect of TOD normalization and $1/f$ filtering for a single feed and scan. The raw Level~1 data (left) are dominated by frequency-dependent gain variations, that correspond to the instrumental bandpass. After normalization (middle), the signal is dominated by common-mode gain fluctuations. Finally, after the $1/f$ filter is applied (right), the common-mode $1/f$ contribution has been suppressed, and the data are dominated by white noise. The horizontal gray stripes indicate channels that were masked by the pipeline. All three stages happen before absolute calibration, and the amplitudes are therefore given in arbitrary units.}
    \label{fig:l2gen_2D_TODs}
\end{figure*}

\subsection{Normalization}\label{sec:l2gen:norm}
The first filtering step in the TOD pipeline is to normalize the Level 1 data by dividing by a low-pass filtered version of the data and subtracting 1. The filter can be written as
\begin{equation}\label{eqn:norm_filter}
    \begin{aligned}
        d^\mathrm{norm}_{\nu, t} = \frac{d^\mathrm{raw}_{\nu, t}}{\expval{d^\mathrm{raw}}_\nu} - 1, \quad \expval{d^\mathrm{raw}}_\nu = \mathcal{F}^{-1}\{W\cdot \mathcal{F}\{{d^\mathrm{raw}_{\nu,t}}\}\}, \\ W = \qty(1 + \qty(\frac{f}{f_\mathrm{knee}})^\alpha)^{-1},
    \end{aligned}
\end{equation}
where $\mathcal{F}$ is the Fourier transform and $\mathcal{F}^{-1}$ is the inverse Fourier transform, both performed on the time-dimension of the data, and $W$ is a low pass filter in the Fourier domain, with a spectral slope $\alpha = -4$, and a knee frequency $f_\mathrm{knee} = \SI{0.01}{Hz}$. This has the effect of removing all modes on $\SI{100}{s}$ timescales and longer. As the COMAP telescope crosses the entire field in 5--20 seconds, the normalization has minimal impact on the sky signal in the scanning direction, but heavily suppresses the signal perpendicular to the scanning direction, as the fields take 5--7 minutes to drift across.

The filter is performed per frequency channel, and the primary purpose of the normalization is to remove the channel-to-channel variations in the gain, making channels more comparable. After applying this filter, the white noise level in each channel becomes the same, and common-mode $1/f$ gain fluctuations become flat across frequency. As a secondary consideration, the normalization also removes the most slowly varying atmospheric and gain fluctuations, on timescales greater than $\SI{100}{s}$.

The implications of the filter become more obvious by stating an explicit data model. The raw detector signal can be modeled by the radiometer equation as a product of gain and brightness temperature, $d^\mathrm{raw} = g_{\nu,t}T_{\nu,t}$, where both terms can in theory vary freely. In practice, the LNA gain can be decomposed into a mean time-independent part, and a multiplicative fluctuation around this mean, such that we get $g_{\nu,t} = g_\nu(1 + \delta g_t)$, where $\delta g_t$ is a small, frequency-independent term (often referred to as $1/f$ gain noise). Because all frequencies are processed by the same LNAs, these fluctuations become common-mode.

We can similarly decompose the brightness temperature into a mean and fluctuation part. However, we can make no assumption about the frequency-dependence of the fluctuation term, and therefore simply write $T_{\nu,t} = T_\nu + \delta T_{\nu,t}$. Here, $T_\nu$ is now the system temperature, as described in Eqn. \ref{eqn:tsys}, and $\delta T_{\nu,t}$ are comparatively small fluctuations around this temperature. Putting this all together we get the data model
\begin{equation}\label{eqn:norm_data_model}
\begin{aligned}
    d^\mathrm{raw} = g_{\nu,t}T_{\nu,t} &= g_\nu\qty(1 + \delta g_t)(T_\nu + \delta T_t) \\
    &= g_\nu T_\nu\qty(1 + \delta g_t + \delta T_{nu,t}/T_\nu + \delta g_t\delta T_{\nu,t}/T_\nu) \\
    &\approx g_\nu T_\nu\qty(1 + \delta g_t + \delta T_{\nu,t}/T_\nu),
\end{aligned}
\end{equation}
where we have used the assumption that both fluctuations terms are small to approximate $\delta g_t\delta T_{\nu,t}/T_\nu \approx 0$.

Under this data model, the low-pass filtered data is assumed to take the form $\langle d^\mathrm{raw}\rangle_\nu = g_\nu T_\nu$, such that the normalization filter has the effect of transforming the data into normalized fluctuations in gain and temperature around zero. Inserting Eqn. \ref{eqn:norm_data_model} into the filter as defined by Eqn. \ref{eqn:norm_filter}, we get the normalized data
\begin{equation}\label{eqn:norm_model_result}
    d^\mathrm{norm}_{\nu, t} = \frac{g_\nu T_\nu\qty(1 + \delta g_t + \delta T_{\nu,t}/T_\nu)}{g_\nu T_{\nu,t}} - 1 = \delta g_t + \delta T_{\nu,t}/T_\nu.
\end{equation}
Technically, we defined $T_\nu$ and $g_\nu$ to have no time-dependence, but in the context of this filter, it makes sense to consider them the temperature and gain terms that fluctuate more slowly than $\SI{\sim 100}{s}$ timescales, as these are the timescales subtracted by this filter.

The effect of the filter can be seen in the first two panels of Fig.~\ref{fig:l2gen_2D_TODs}, which shows the TOD of a single scan in 2D before and after the normalization. Before the normalization, the frequency-dependent gain dominates, and the time variations are invisible. After normalization, the data in each channel fluctuates around zero.

\subsection{Azimuth filter}
Next, we fit and subtract a linear function in azimuth, to reduce the impact of pointing-correlated systematic errors, first and foremost being ground pickup by the telescope sidelobes. This filter can be written as
\begin{equation}
    d^\mathrm{az}_{\nu, t} = d^\mathrm{norm}_{\nu, t} - a_\nu\cdot \mathrm{az}_t,
\end{equation}
where $a_\nu$ is fitted to the data per frequency, and $\mathrm{az}_t$ is the azimuth pointing of the telescope. Unlike in the ES pipeline, this filter is now fitted independently for when the telescope is moving eastward and westward, to mitigate some directional systematic effects we have seen.

In Season 1, we also employed Lissajous scans, meaning that an elevation term was also present in this equation. As we now only observe in constant elevation mode, this term falls away.

\subsection{$1/f$ gain fluctuation filter}
\label{sec:l2gen:polyfilter}
After normalization, the data are dominated primarily by gain, and secondarily by atmospheric fluctuations, and both are strongly correlated on longer timescales. Although the normalization suppresses power on all timescales longer than 100 seconds, we observe that common-mode noise still dominates the total noise budget down to ${\sim}\SI{1}{s}$ timescales.

To suppress this correlated noise, we apply a specific $1/f$ filter\footnote{The filter is referred to as the polynomial filter in our ES publications.} by exploiting the simple frequency behavior of the gain and atmosphere fluctuations. After we have normalized the data, the amplitude of the gain fluctuations is the same across all frequency channels, although fluctuating in time, as we see from the term $\delta g_t$ in Eqn. \ref{eqn:norm_model_result}. The atmospheric fluctuations will enter the $\delta T_{\nu,t}$ term of Eqn. \ref{eqn:norm_model_result}. To model the atmosphere, we approximate both the atmosphere and the system temperature $T_\nu$ as linear functions of frequency, such that the combined term $\delta T_{\nu,t}/T_\nu$ is itself linear. To jointly remove the gain and atmosphere fluctuations, we therefore fitted and subtracted a first-order polynomial across frequency for every time step:
\begin{equation}
    d^\mathrm{1/f}_{\nu, t} = d^\mathrm{point}_{\nu, t} - (c^0_t + c^1_t \cdot \nu),
\end{equation}
where $c_t^0$ and $c_t^1$ are coefficients fitted to the data each time step. To ensure that both the atmosphere and the system temperature are reasonably well approximated as linear in frequency, we fit a separate linear polynomial to the 1024 channels of each of the four $\SI{2}{GHz}$ sidebands.

This simple technique is remarkably efficient at removing $1/f$ noise, and we observe that the correlated noise is suppressed by several orders of magnitude, after which white noise dominates the uncertainty budget. This is illustrated in the last two panels of Fig.~\ref{fig:l2gen_2D_TODs}. After the normalization, the signal is completely dominated by common-mode $1/f$ noise. After the $1/f$ filter, the correlated noise is effectively suppressed, and we are left with almost pure white noise, as can be seen in the right panel, and further discussed in Sect.~\ref{sec:TOD-results}.

\subsection{PCA filtering}
\label{sec:TOD-PCA}
Principal component analysis (PCA) is a common and powerful technique for dimensionality reduction \citep{doi:10.1080/14786440109462720}. Given a data matrix $m_{\nu,t}$, PCA produces an ordered basis $w_t^k$ for the columns of $m_{\nu,t}$, called the principal components of $m_{\nu,t}$. The component amplitude can then be calculated by re-projecting the components into the matrix, as $a^{k}_\nu = m_{\nu,t} \cdot w^k_t$. For our purposes, $m_{\nu,t}$ is the TOD, with frequencies as rows, and time-samples as columns. The ordering of the principal components $w_t^k$ is such that the earlier components capture as much of the variance in the columns of $m_{\nu,t}$ as possible, and for any selected number of components $N_\mathrm{comp}$, the following expression is minimized:
\begin{equation}
    \sum_{\nu,t}\qty(m_{\nu,t} - \sum_{k=1}^{N_\mathrm{comp}}a_\nu^k w_t^k)^2.
\end{equation}
In other words, PCA provides a compressed version of $m_{\nu,t}$, that approximates $m_{\nu,t}$ as the sum of an ordered set of outer products\footnote{PCA has several equivalent interpretations and ways of solving for the principal components. The principal components are, among other things, the eigenvectors of the covariance matrix of $m_{\nu,t}$. This is how we introduced the PCA in our ES publications. It is, however, both a slow way of solving for the PCA components in practice and not the best interpretation for our purposes.}
\begin{equation}
    m_{\nu,t} \approx \sum_{k=1}^{N_\mathrm{comp}}a_\nu^k w_t^k.
\end{equation}

PCA is often employed on a dataset where the rows are interpreted as different observations, and the columns are the multi-dimensional features of these data. However, this is not a natural interpretation for our purposes, and it makes more sense to simply look at PCA as a way of compressing a 2D matrix as a sum of outer products -- we have no special distinction between columns and rows, and could equivalently have solved for the PCA of the transpose of $m_{\nu,t}$, which would swap $a_\nu$ and $w_t$.

A PCA is often performed because one is interested in keeping the leading components, as these contain much of the information in the data. However, we subtract the leading components, because many systematic errors naturally decompose well into an outer product of a frequency vector and a time vector, while the CO signal does not (and is very weak in a single scan).

In practice, we solve for the leading principal components using a singular value decomposition algorithm \citep{doi:10.1137/090771806}, and then calculate the amplitudes as stated above. The $N_\mathrm{comp}$ leading components are then subtracted from the TODs, leaving us with the filtered data
\begin{equation}
    d^\mathrm{PCA}_{\nu,t} = m_{\nu,t} - \sum_{k=1}^{N_\mathrm{comp}} a^k_\nu w^k_t.
\end{equation}

The Season 2 COMAP pipeline employs two time-domain PCA filters, one of which was present in ES. In the following subsection, we introduce both filters and then explain how to decide the number of leading components, $N_\mathrm{comp}$, to subtract from the data.

\begin{figure}
    \centering
    \includegraphics[width=\linewidth]{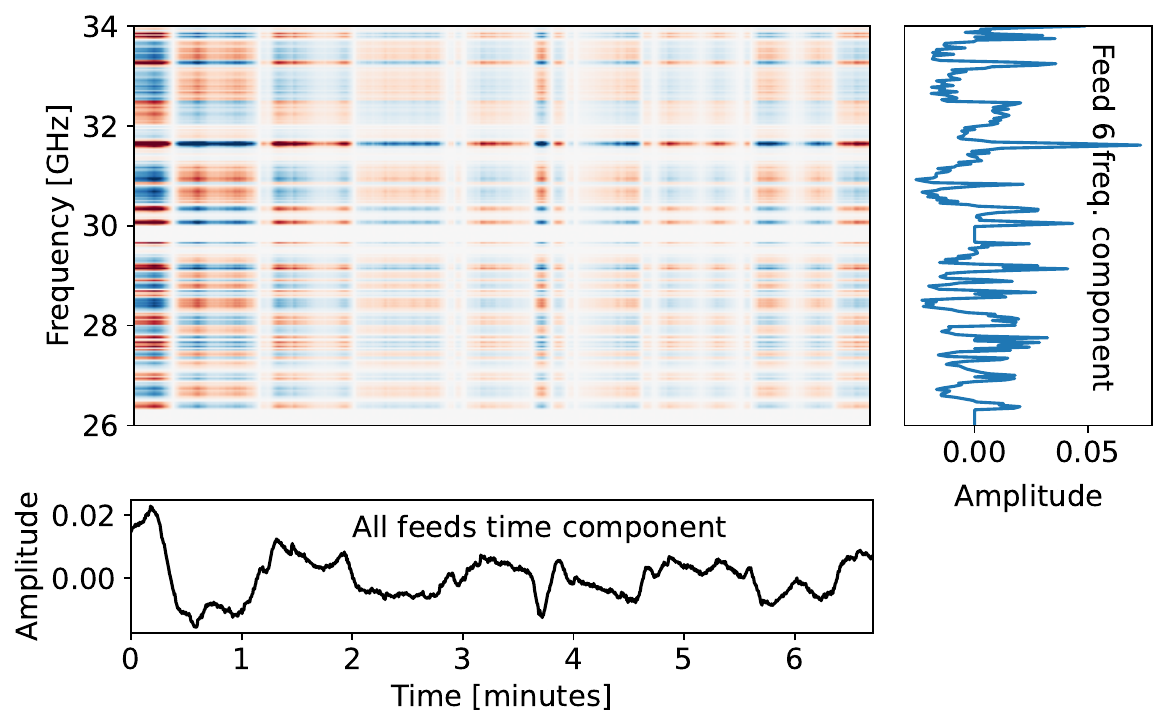}
    \vspace{-4mm}\caption{Most significant component and amplitude of the all-feed PCA filter applied on scan 3354205. The bottom plot shows the first PCA component $w_t^0$, which is common to all feeds. The right plot shows the corresponding amplitude $a_\nu^0$ for feed 6. The outer product of these two plots is shown in the central image and is the quantity subtracted from the TOD by the filter. As the filter is applied on normalized data, the amplitudes are all unitless, and the colorbar limits are $\pm 5\times 10^{-4}$, half the range of those in Fig \ref{fig:l2gen_2D_TODs}.}
    \label{fig:comp_ampl_aPCA}
\end{figure}

Figure \ref{fig:comp_ampl_aPCA} shows an example of a strong component $w_t$ picked up by this filter in the bottom panel, and the corresponding component amplitude $a_\nu$ for feed 6 in the right panel. The center image shows the resulting full frequency- and time-dependent outer product between the two functions, which is the quantity subtracted from the TOD. The time-dependent component is in this example strongly temporally correlated, and could be residual atmospheric fluctuations, that typically have $1/f$-behavior with a steep frequency slope.

The process of calculating the principal components and subtracting them from the data constitutes a non-linear operation on the data. This has the advantage of being much more versatile against systematic errors that are difficult to model using linear filters, but the disadvantage is a more complicated impact on the CO signal itself. This is further discussed in Sect.~\ref{sec:pca_linearity}, where our analysis shows that the PCA filter behaves linearly with respect to any sufficiently weak signal, and that, at the scan-level, all plausible CO models \citep{Chung_2024} are sufficiently weak by several orders of magnitude.

\subsubsection{The all-feed PCA filter}
The all-feed PCA filter, which was also present in the ES pipeline, collapses the 19 feeds onto the frequency axis of the matrix, producing a data matrix $m_{\nu,t}$ of the $1/f$-filtered data with a shape of $(N_\mathrm{feed}~N_\mathrm{freq}, N_\mathrm{TOD}) = (19\times 4096, {\sim}20\,000)$ for a scan with $N_\mathrm{feed}$ feeds, $N_\mathrm{freq}$ frequency channels, and $N_\mathrm{TOD}$ time samples. The PCA algorithm outlined above is then performed on this matrix. Combining the feed and frequency dimensions means that a feature in the data will primarily only be picked up by the filter if it is common (in the TOD) across all 19 feeds. This is primarily the case for any atmospheric contributions, and potentially standing waves that originate from the optics common to all feeds. It is, however, certainly not the case for the CO signal, which will be virtually unaffected by this filter.    

\subsubsection{The per-feed PCA filter}
The new per-feed PCA filter has been implemented to combat systematic errors that vary from feed to feed. This filter employs the PCA algorithm outlined above on each individual feed and is performed on the output of the all-feed PCA filter. Additionally, we found that downsampling the data matrix (using inverse variance noise weighing) by a factor of 16 in the frequency direction before performing the PCA increased its ability to pick up structures in the data. The resulting data matrix $m_{\nu,t}$ gets the shape $(N_\mathrm{freq}/16, N_\mathrm{TOD}) = (256, {\sim}20\,000)$ for each feed. The downsampling is only used when solving for the time-domain components $w_t$, and the full data matrix is used when calculating the frequency amplitudes, $a_\nu$.

Targeting each feed individually makes us more susceptible to CO signal loss, but the low signal-to-noise ratio (S/N), combined with the fact that the CO signal cannot be naturally decomposed into an outer product, makes the impact on the CO signal itself minimal. This filter appears to primarily remove components consistent with standing waves from the individual optics and electronics of each feed.

\begin{figure}
    \centering
    \includegraphics[width=\linewidth]{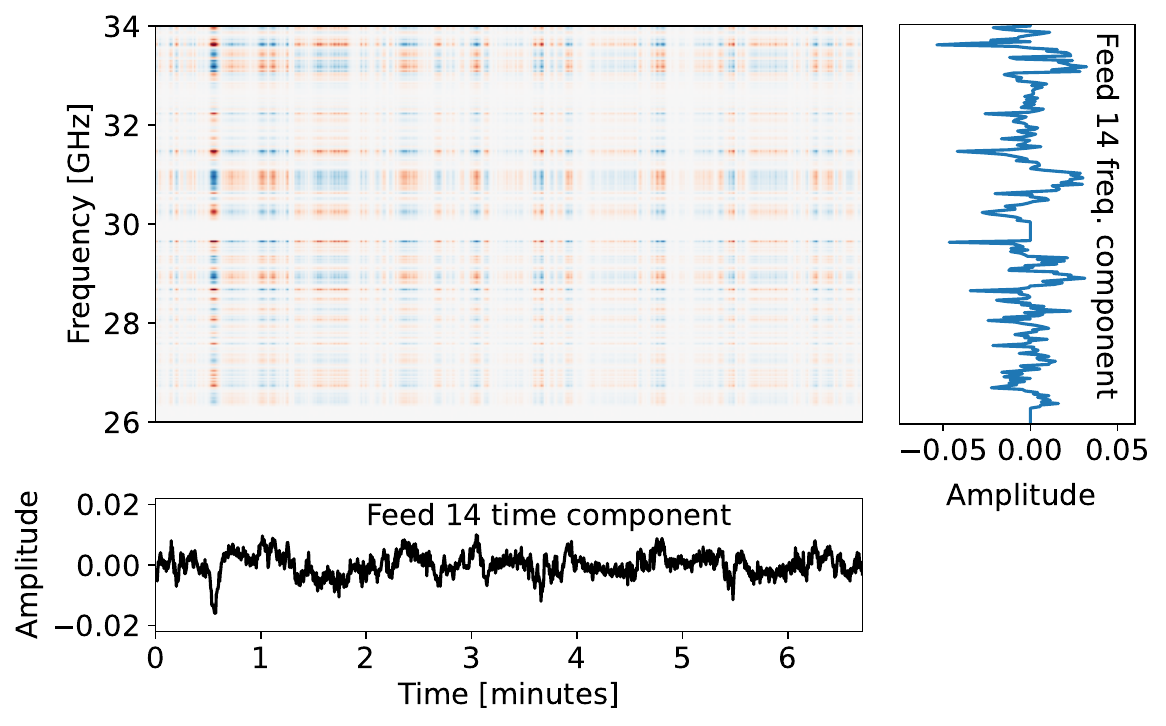}
    \vspace{-4mm}\caption{Most significant component and amplitude of the per-feed PCA filter applied on scan 3354205. The bottom plot shows the first PCA component $w_i^0$ for feed 14, while the right plot shows the corresponding amplitude $a_i^0$ for feed 14. The other product of the two quantities are shown in the center plot, with colorbar limits of $\pm 5\times 10^{-4}$.}
    \label{fig:comp_ampl_fPCA}
\end{figure}

Figure \ref{fig:comp_ampl_fPCA} shows a typical strong component picked up by the per-feed PCA filter for feed 14, similar to what was shown for the all-feed PCA filter in Fig. \ref{fig:comp_ampl_aPCA}. The per-feed PCA filter typically picks up somewhat weaker features than the all-feed PCA filter, as it is applied after the all-feed PCA filter. The time-dependent component in Figure \ref{fig:comp_ampl_fPCA} is noticably noisier than the equivalent component \ref{fig:comp_ampl_aPCA}, also owing to the fact that it is fit with the data from 1 feed, as opposed to 19 for the all-feed filter. The physical origin of the feature shown is unknown.

\subsubsection{Number of components}
In the ES pipeline, the number of PCA components was fixed at four for the all-feed filter, and the per-feed filter did not exist. We now dynamically determine the required number of components for each filter, per scan. This allows us to use more components when needed, removing more systematic errors, and fewer when not needed, incurring a smaller loss of CO signal.

We subtract principal components until the components are indistinguishable from white noise, that can be inferred from the singular values of each component. Let $\lambda$ be the expectation value of the largest singular value of a $(N, P)$ Gaussian noise matrix (see Appendix \ref{app:pca_threshold} for how this value is derived). We subtract principal components until we reach a singular value below $\lambda$. However, for safe measure, we always subtract a minimum of 2 components for the all-feed PCA filter, as the signal impact of this filter is minimal.

\begin{figure}
    \centering
    \includegraphics[width=\linewidth]{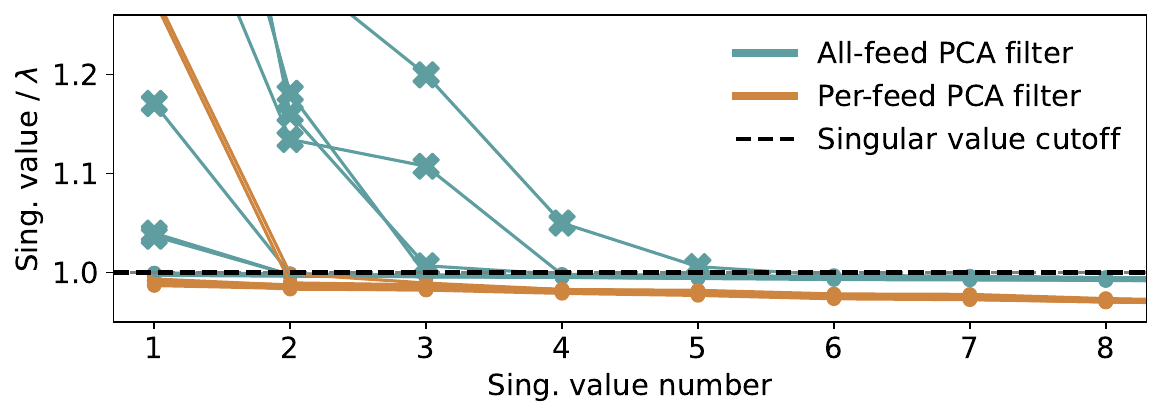}
    \includegraphics[width=\linewidth]{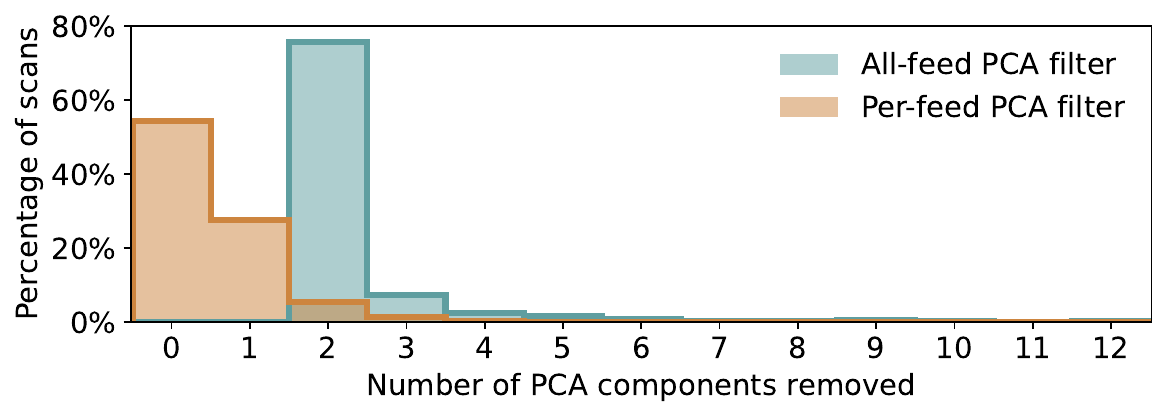}
    \vspace{-4mm}\caption{\emph{(Top:)} Largest singular values of the all-feed and per-feed PCA filters, divided by $\lambda$, for a random selection of ten scans. PCA components with relative values above one are removed from the data and are marked with crosses in this plot. \emph{(Bottom:)} Number of PCA components subtracted across all scans. At least two components are always subtracted by the all-feed PCA filter.}
    \label{fig:num_PCA}
\end{figure}

Figure \ref{fig:num_PCA} shows typical singular values, relative to $\lambda$, for a random selection of scans, and a histogram of the number of components employed across all scans. The average number of PCA components subtracted is 2.3 and 0.5 for the all-feed and per-feed PCA, respectively; the most common number of components subtracted is the minimum allowed in each case: two and zero. The top part of the figure also demonstrates that there is a sharp transition between the singular values of the components that actually pick up meaningful features from the signal, and the remaining noise components. For reference, the quite significant components shown in Fig. \ref{fig:comp_ampl_aPCA} and Fig. \ref{fig:comp_ampl_fPCA} had singular value to $\lambda$ ratios of 2.9 and 1.6, respectively.

\subsection{Data-inferred frequency masking}
\label{sec:l2gen:masking}
After the PCA filters, we perform dynamic masking of frequency channels identified from the filtered TOD. This mainly consists of masking groups of channels that have substantially higher correlations between each other than expected from white noise, explained in more detail in \cite{Foss_2022}. This is assumed to be caused by substantial residuals of gain fluctuations or atmospheric signal. We did this by calculating the correlation matrix between the frequency channels and perform $\chi^2$ tests on the white noise consistency of boxes and stripes of various sizes across this matrix. We also mask individual channels with a standard deviation significantly higher than expected from the radiometer equation.

After the frequency masking, the $1/f$ filter and PCA filters are reapplied to the data, to ensure that their performance was not degraded by misbehaving channels. The normalization and pointing filters do not need to be reapplied, as they work independently on each frequency channel.

\subsection{Calibration and downsampling}\label{sec:l2gen:calib}
The final step of the pipeline is to calibrate the data to temperature units and decrease the frequency resolution. After the normalization step, the data are in arbitrary normalized units. Using the system temperature calculated in Sect. \ref{sec:l2gen:tsys_calc} we calibrate each channel of the data,
\begin{equation}
    d^\mathrm{cal}_{\nu, t} = T^\mathrm{sys}_\nu d^\mathrm{PCA}_{\nu, t}.
\end{equation}
Finally, we downsample the frequency channels from 4096 native channels to 256 science channels. The Seasons 1 and 2a frequency channels are $\SI{1.953}{MHz}$ wide, while they are $\SI{2.075}{MHz}$ wide in Season 2b, after the change in sampling frequency (Sect.~\ref{sec:aliasing}). In both cases, the channels are downsampled to a grid of $\SI{31.25}{MHz}$, that exactly corresponds to a factor 16 downsampling for the older data. For the newer data, either 15 or 16 native channels will contribute to each science channel, decided by their center frequency. The downsampling is performed with inverse-variance weighting, using the system temperature as uncertainty.

\subsection{Implementation and performance}
\label{sec:l2gen_perf}
While the ES pipeline was written in Fortran, the Season 2 pipeline has been rewritten from scratch to run in Python. Performance-critical sections are either written in C++ and executed using the Ctypes package or employ optimized Python packages such as SciPy. Overall, the serial performance is similar to the ES pipeline, but the Season 2 pipeline employs a more fine-grained and optimal MPI+OpenMP parallelization, making it much faster on systems without a very large memory-to-core ratio.

The pipeline is run on a small local cluster of 16 E7-8870v3 CPUs, with a total of 288 cores, in about a week of wall time, totaling around 40,000 CPU hours for the full COMAP dataset. The time-domain processing dominates this runtime, with a typical scan taking around 20--25 minutes to process on a single CPU core.

\subsection{Time-domain results}\label{sec:TOD-results}
\begin{figure}
    \centering
    \includegraphics[width=\linewidth]{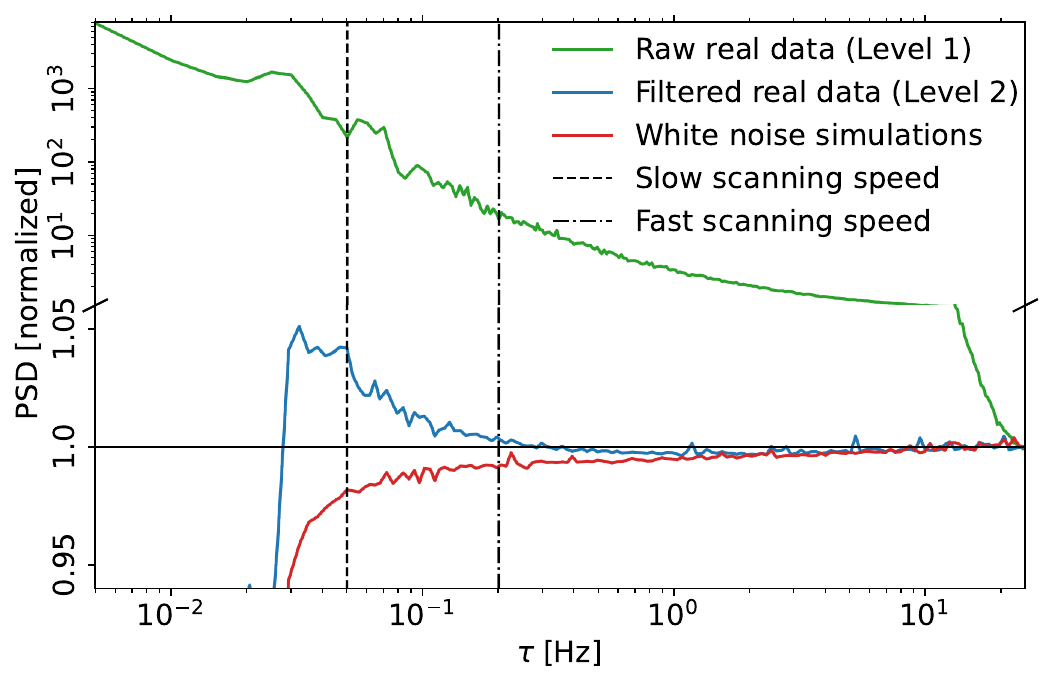}
    \vspace{-4mm}\caption{Average temporal power spectrum of unfiltered (green) and filtered scans (blue), compared to correspondingly filtered white noise simulations (red). The y-axis is broken at 1.06 and is logarithmic above this. The data are averaged over scans, feeds, and frequencies, and are normalized with respect to the highest $k$-bin. For context, the old and new scanning frequencies (once across the field) are shown as vertical lines (dot-dashed and dashed, respectively).
    }
    \label{fig:TOD-PS}
\end{figure}

The Level 2 TODs outputted by \texttt{l2gen} are assumed to be almost completely uncorrelated in both time and frequency dimensions, such that the TOD are well approximated as white noise. To quantify the correlations in the time domain we calculate the temporal power spectrum of each individual channel for all scans. Figure \ref{fig:TOD-PS} shows this power spectrum averaged over both scans and frequencies, compared both to the equivalent power spectrum of un-filtered Level 1 data, and to that of a TOD obtained by injecting pure white noise in place of our real data into the TOD pipeline.

Since the pipeline filtering removes more data on longer timescales, the white noise simulation (red) gradually deviates from a flat power spectrum on longer timescales, until it falls rapidly at timescales below $\sim \SI{0.03}{Hz}$ due to the high-pass normalization performed in the pipeline. The power spectrum for the filtered real data (blue) also follows the same trend on short timescales, but then increases on timescales around $k=\SI{0.2}{Hz}$, due to small residual $1/f$ gain fluctuations and atmosphere remaining after the processing. Below $\sim \SI{0.03}{Hz}$, this power spectrum again falls rapidly due to the high-pass filter. The power spectrum obtained from raw data that have not gone through any filtering (green), simply increases on longer timescales as expected, due to $1/f$ gain and atmospheric fluctuations.

The difference between the blue and red spectra shows the residual correlated noise left in the data. While there is some residual correlated noise, it is an insignificant fraction of our final noise budget. We find that our real filtered data have a standard deviation only 1.7\,\% higher than that of the filtered white noise. Compared to the amount of power we see in the unfiltered data, we see that our pipeline is very efficient at suppressing correlated noise.

\section{Mapmaking and map domain filtering}
\label{sec:mapmaking}

\subsection{The COMAP mapmaker}
\label{sec:mapmaker}
COMAP employs a simple binned inverse-variance noise-weighted mapmaker, identical to the one in ES \citep{Foss_2022}. This can be written as
\begin{equation}
    m_{\nu,p} = \frac{\sum_{t\in p}d_{\nu,p}/\sigma_{\nu,t}^2}{\sum_{t\in p}1/\sigma_{\nu,t}^2},
\end{equation}
where $m_{\nu,p}$ is an individual map voxel\footnote{A voxel here is the 3D equivalent of a pixel, with two angular dimensions and a redshift (frequency) dimension. Here we separate the voxel dimensions into frequencies $\nu$ and pixels $p$.}, $d_{\nu,t}$ represents the time-domain data over time samples, $\sigma_{\nu,t}^2$ is the time-domain white noise uncertainty, and $t\in p$ means all time-samples $t$ that observes pixel $p$. We assume that the white noise uncertainty $\sigma_{\nu,t}$ is constant (for a single feed and frequency) over the duration of a scan, and calculate it per scan as
\begin{equation}
    \sigma_\nu = \sqrt{\frac{\mathrm{Var}(d_{\nu,t} - d_{\nu,t-1})}{2}},
\end{equation}
then let $\sigma_{\nu,t} = \sigma_{\nu}$ for all time-samples within the scan. This value is also binned into maps, and is used as the uncertainty estimate of the maps throughout the rest of the analysis:
\begin{equation}\label{eqn:map_sigma}
    \sigma_{\nu,p}^2 = \frac{1}{\sum_{t\in p}1/\sigma_t^2}.
\end{equation}
In practice, we calculate per-feed maps, both because the map-PCA filter (introduced in the next section) is performed on per-feed maps, and because the cross-spectrum algorithm utilizes groups of feed-maps.

The reason for not using more sophisticated mapmaking schemes, such as destriping \citep{Keih_nen_2010} or maximum likelihood mapmaking \citep{Tegmark_1997}, is partially of necessity -- the COMAP TOD dataset is many hundreds of TB, making iterative algorithms difficult. However, the TOD pipeline has also proven remarkably capable of cleaning most unwanted systematic errors from the data, especially correlated $1/f$ noise, as we saw from Fig.~\ref{fig:TOD-PS}. As the main purpose of more sophisticated mapmaking techniques is dealing with correlated noise, COMAP is served well with a simple binned mapmaker.

The mapmaking algorithm is identical to the one used for the ES analysis. However, as with \texttt{l2gen} the actual implementation has been rewritten from scratch in Python and C++, with a focus on optimal parallelization and utilization of both MPI and OpenMP.

\subsection{Map-domain PCA filtering}\label{sec:mPCA}
The pipeline now employs a PCA filtering step also in the map domain, in addition to the one we apply at the TOD level. This technique is almost entirely analogous to the PCA foreground subtraction often employed in 21cm LIM experiments \citep{2010Natur.466..463C,Masui_2013,Anderson_2018}, although we did not employ it to subtract foregrounds. The primary purpose of this filter is to mitigate a couple of pointing-correlated systematic errors (see the next subsection) which proved challenging to remove entirely in the time domain. The method is similar to the TOD PCA algorithm from Sect.~\ref{sec:TOD-PCA}, but instead of having the TOD data matrix $d_{\nu,t}$, we have a map $m_{\nu,p}$ with one frequency and one (flattened) pixel dimension. The data matrix then gets the shape $(N_\nu, N_p) = (256,\, 14400)$, with $N_\nu=256$ being the number of frequency channels in the map and $N_p=120\cdot 120 = 14400$ the number of pixels in each frequency slice (although many of the pixels in each individual feed-map are never observed).

The technique we employ here is technically a slight generalization of the PCA problem, as we want to weigh individual voxels by their uncertainty when solving for the components and amplitudes\footnote{For the time-domain PCA, it was enough to weight individual channels, with all time-samples in that channel sharing the same weight. This can be done with a normal PCA, as we show in Appendix \ref{app:mpca_solution}.}. This is not possible in the regular PCA framework without also morphing the modes one is trying to fit, as we explain in Appendix \ref{app:mpca_solution}. As shown in Sect. \ref{sec:TOD-PCA}, the first principal component $w^0_p$ and its amplitude $a^0_\nu$ are the vectors that minimize the value of the expression
\begin{equation}
    \sum_{\nu,p} (m_{\nu,p} - a^0_\nu w^0_p)^2.
\end{equation}

In other words, they are the two vectors such that their outer product explains as much of the variance in $m_{\nu,p}$ as possible. This formulation of the PCA makes it obvious how to generalize the problem to include weighting for individual matrix elements: we can simply minimize the following sum,
\begin{equation}
    \label{eqn:mPCA1}
    \sum_{\nu,p} \frac{(m_{\nu,p} - a^0_\nu w^0_p)^2}{\sigma_{\nu,p}^2},
\end{equation}
where $\sigma_{\nu, p}$ is the uncertainty in each voxel. Minimizing Eq. (\ref{eqn:mPCA1}) gives us the vectors $a^0_\nu$ and $w^0_p$ for which the outer product $a^0_\nu w^0_p$ explains as much of the variance in $d_{\nu,p}$ as possible, weighted by $\sigma_{\nu,p}$. The resulting map $m_{\nu,p} - a^0_\nu w^0_p$ represents the filtered map. The process can then be repeated any number of times, solving for and subtracting a new set of vectors. We minimize the expression in Eq.~(\ref{eqn:mPCA1}) iteratively with an algorithm outlined in Appendix \ref{app:mpca_solution}, where we also explain why this is not equivalent to simply performing the PCA on a noise-weighted map $m_{\nu,p}/\sigma_{\nu,p}$. Due to the large similarity of our technique to a regular PCA, we simply refer to this filter as a PCA filter.

As for the TOD PCA filter, a selected number of components are subtracted from the data maps 
\begin{equation}
    m^\mathrm{mPCA}_{\nu,p} = m_{\nu,p} - \sum_{i=1}^{N_\mathrm{comp}}a^k_\nu v^k_p,
\end{equation}
This filtering is performed per feed, as the systematic errors outlined in the next subsection manifest differently in different feeds. We have chosen $N_\mathrm{comp}=5$, which is further explained in Sect. \ref{sec:effect_mpca}.  Because the COMAP scanning strategy stayed the same throughout Seasons 1 and 2a but changed with the azimuth slowdown of Season 2b, we apply the map-PCA separately to the former and latter, as the pointing-correlated effects we are trying to remove might also be different.

\subsection{Newly discovered systematic effects}
\begin{figure}
    \centering
    \includegraphics[width=\linewidth]{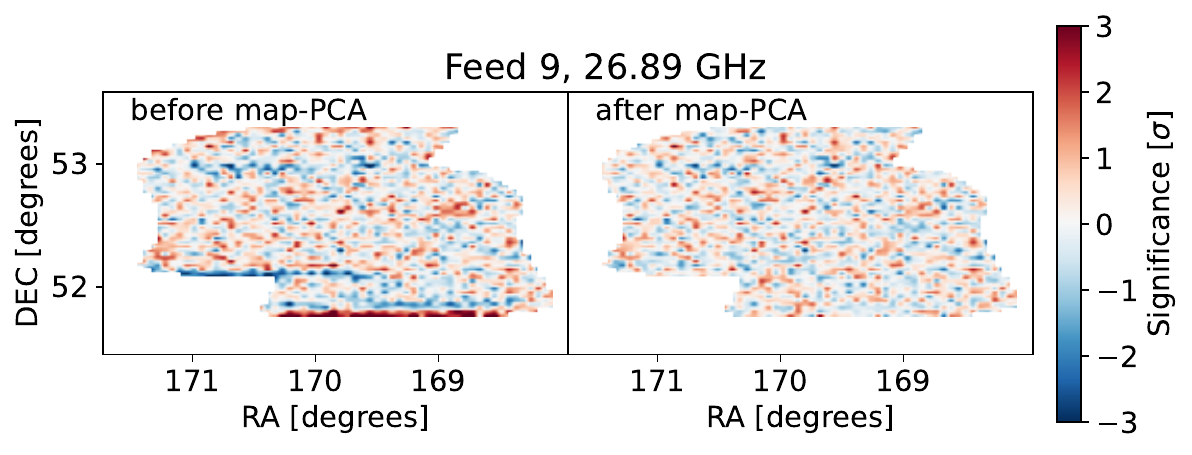}
    \includegraphics[width=\linewidth]{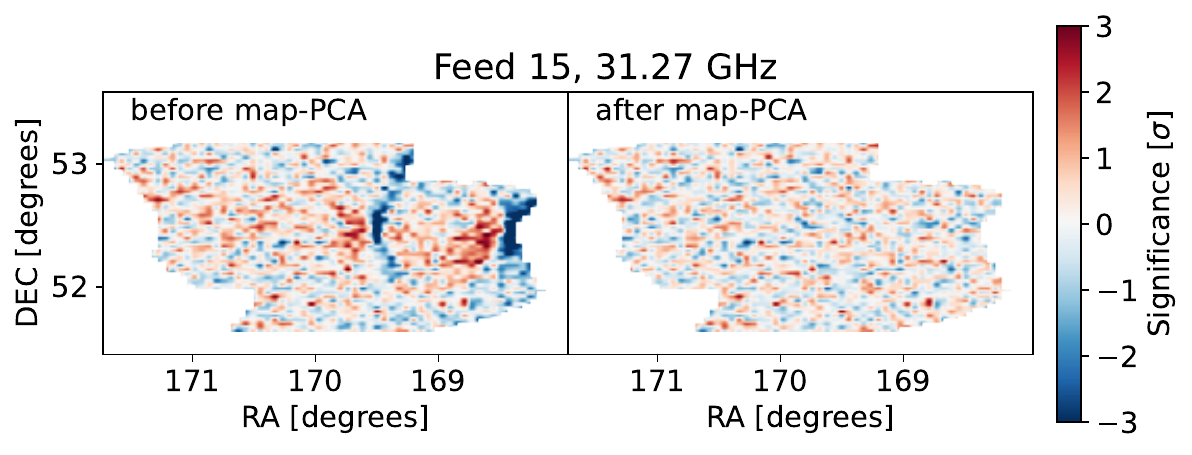}
    \includegraphics[width=\linewidth]{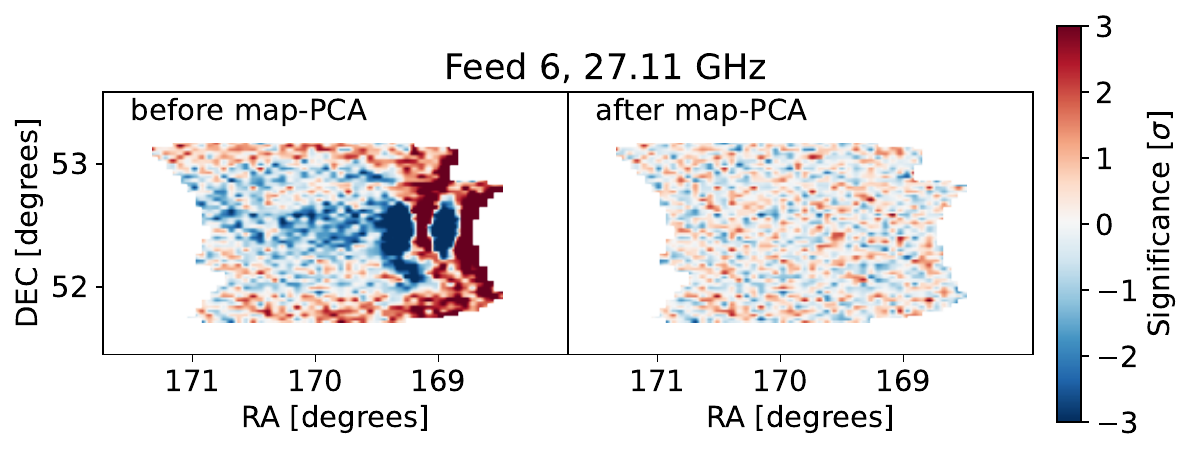}
    \includegraphics[width=\linewidth]{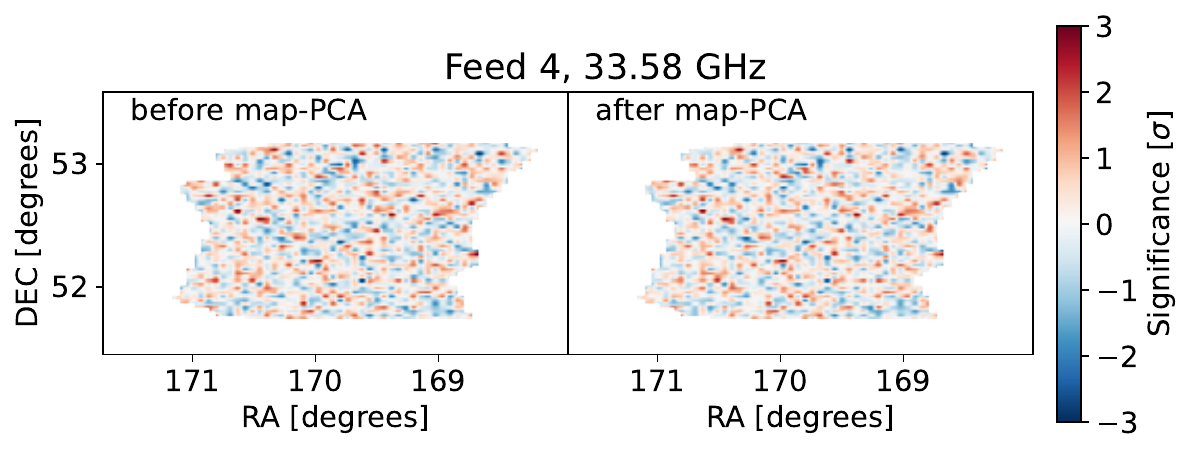}
    \vspace{-4mm}\caption{Selection of individual frequency maps from Field 2 before \emph{(left)} and after \emph{(right)} the map-domain PCA filter. Because of the uneven sensitivity among the pixels, and to emphasize the relevant systematic effects, all maps have been divided by their white noise uncertainty. From top to bottom, each row shows maps that are dominated by (1) the turn-around effect; (2) the start-of-scan effect; (3) both effects simultaneously; and (4) neither effect. Both effects appear to manifest twice, on two slightly offset maps. This offset effect originates from the physical placement of the feeds, as they observe the fields as they are both rising and setting on the sky. In the equatorial coordinate system of these maps, the telescope scans vertically, and the field drifts from left to right.
    }
    \label{fig:mPCA_figures}
\end{figure}
\begin{figure}
    \centering
    \includegraphics[width=\linewidth]{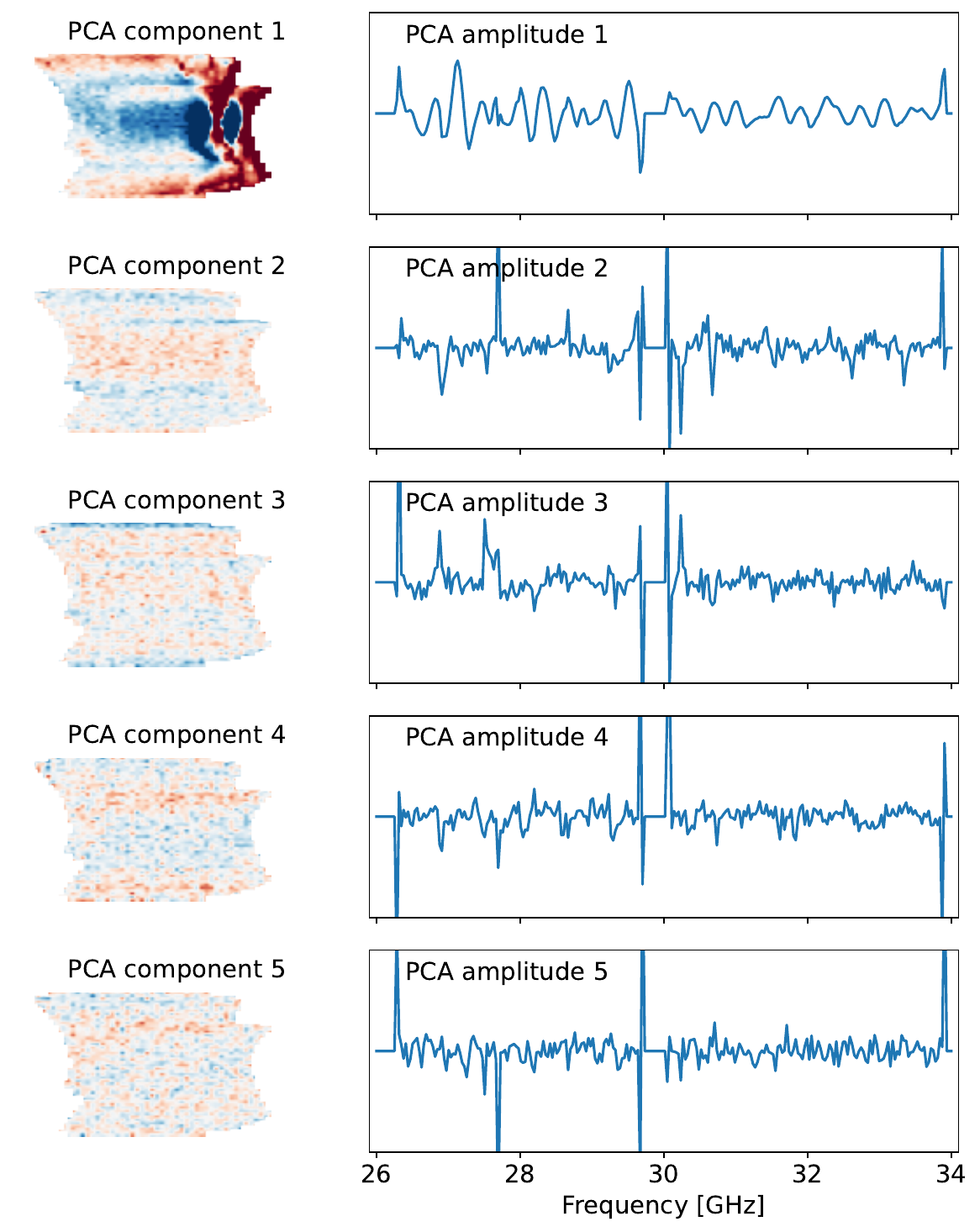}
    \vspace{-4mm}\caption{Leading PCA components $v^k$ \emph{(left)} and their respective frequency amplitudes $a^k$ \emph{(right)} for Field 2 as observed by feed 6 prior to the map-PCA filter; this feed is the most sensitive to pointing-correlated systematics.
    All maps are divided by their respective uncertainties to highlight the key morphology. All rows share the same color range and $y$-axis scale, but the specific values have been omitted as they are not easily interpretable.}
    \label{fig:mPCA_components}
\end{figure}
\label{sec:new-systematics}
The two most prominent new systematic errors discovered in the second season of observations have been dubbed the ``turn-around'' and ``start-of-scan'' effects. They have in common that they are difficult to model in the time domain, subtle in individual scans, but strongly pointing-correlated, and they therefore show up as large-scale features in the final maps. Additionally, they are present to varying extents in all feeds, have similar quantitative behavior in the map-domain, and can both be removed effectively with the map-PCA. The effects are discussed in the subsections below, with further analysis shown in Appendix~\ref{app:new-systematics}.

\subsubsection{The turn-around effect}
The so-called ``turn-around'' effect can be observed as strongly coherent excess power near the edges of the scan pattern, where the telescope reverses direction in azimuth. Illustrations of this effect can be seen in the first row, and partially in the third row, of Fig.~\ref{fig:mPCA_figures}. The feature manifests at the top and bottom of the maps, as this is where the telescope turn-around happens for Field 2 in equatorial coordinates. The feature oscillates slowly across the frequency domain, and the leading theory of its origin is some standing wave oscillation induced by mechanical vibrations. The effect is somewhat less pronounced in Season 2b, after the reduction in telescope pointing speed and acceleration, but it is still present.

\subsubsection{The start-of-scan effect}

A related effect is called the ``start-of-scan'' effect, that is a wave-like feature in frequency that occurs at the beginning of every scan and decays exponentially with a mean lifetime of around 19\,sec. As the telescope always starts each scan at the same side of each sky field, this systematic effect shows up in the map domain as a strong feature on the Eastern edge of the map, as can be seen in the second and third rows of Fig.~\ref{fig:mPCA_figures}. Next to the strong positive or negative signal (this varies by frequency) at the very edge of the map, the opposite power, at lower amplitude, can be observed as we move Westward across the map. This opposite power is simply a ringing feature from the normalization performed during the pipeline (see Appendix \ref{app:new-systematics} for details).

The exact origins of the ``start-of-scan'' effect are unknown, but the fact that it only happens at the beginning of scans (that are separated by a repointing to catch up with the field), and disappears during constant elevation scanning, suggests that a potential candidate is mechanical vibrations induced by the repointing. We also observe the effect to be mostly associated with one of the four DCM1s (first downconversion module), namely DCM1-2, relating to feeds 6, 14, 15, 16, and 17. The effect's strong correlation with DCM1-2 points to a possible source in the local oscillator cable, shared by the channels in a DCM1 module; imperfect isolation of the mixer would cause a weak common-mode resonance to manifest.

An important detail in this analysis is that, because of the normalization and $1/f$ filter in the TOD pipeline, any standing wave signal with a constant resonant cavity wavelength over time will be filtered away. For a standing wave to survive the filtering, it must have a changing wavelength. Prime suspects for the origin of this effect are therefore optical cavities that could expand or contract in size, or cables that could be stretched.

\subsubsection{Effects of map-domain filtering}\label{sec:effect_mpca}
The map-domain PCA filtering was implemented in an attempt to mitigate these systematic effects and it has proved to be effective at this task. The first PCA component alone subtracts both effects to a level where they are not visible in the maps by eye. This shows that both effects are well modeled as the outer product of a pixel vector and a frequency vector.

Visually inspecting the PCA components, we usually see some structure for the first 3--5 modes. Figure \ref{fig:mPCA_components} shows an example of the five leading PCA components and their amplitudes. Here we can see that the first component has very clear structure in both the map and frequency domain. The remaining modes seem to absorb some residuals after this first mode, especially on specific channels close to the edges of each of the two Bands that divide the frequency range in two.

We have chosen to remove just five out of 256 PCA components in the map-PCA filter, as no structure was visible by eye in the worst-affected cases after this number of components was removed, and the removal of more components did not significantly affect the results of any subsequent analysis (such as the power spectrum). With only five components being removed, we also limit the potential for CO signal loss. We could have employed a similar approach to the TOD PCA, with a dynamic amount of components, but the noise properties of the maps are more complicated than the TODs, and we have chosen to keep a static number of components, postponing more fine-tuning to future analysis. The filter is applied to the individual feed maps, and to individual splits -- both the elevation split used for the cross power spectrum, and the individual map splits for the null tests.

With the application of the map-PCA filter, we observe that the start-of-scan and turn-around effects are suppressed well below the white noise of the maps, as can be seen in Fig.~\ref{fig:mPCA_figures}. We have also designed a null test to specifically target the turn-around systematic effect, by splitting the maps at the TOD level into east- and west-moving azimuthal directions. The turn-around effect manifests very differently in each half of this split, making it the basis for a sensitive null test. We find that the Season 2 data passes this test after the map-PCA filter has been applied \citep{Stutzer_2024}.

The standard deviation of the maps only falls by 2\% after applying the filter, as the noise still dominates the overall amplitude. However, smoothing the maps slightly will enhance large-scale correlations, while suppressing uncorrelated noise. Smoothing both the filtered and unfiltered 3D maps using a Gaussian with $\sigma=\SI{3}{voxels}$, the standard deviation is 67\% lower in the map-PCA filtered map. We can similarly observe that the average correlation between neighboring pixels (of the unsmoothed map), a good indication of the level of larger scale structure, falls from 6.3\% to $-0.4\%$ after applying the map-PCA.

The magnitude of these systematic effects is different between feeds and frequencies, as seen in Figure \ref{fig:mPCA_figures}. Performing a $\chi^2$ white noise consistency test on the individual frequency channel maps of each feed, we find that for the worst feeds, namely those associated with DCM1-2 (feeds 6, 14, 15, 16 and 17), around 50\,\% of their channels fail this test at $> 5\sigma$. The best-behaving feeds are 4, 5, 10, and 12, all with fewer than 5\,\% of channels failing at $>3\sigma$. However, we want to emphasize both that no feed is completely without these effects before the map-PCA, and that after the map-PCA, there is no longer a quantitative difference between the ``good'' and ``bad'' feeds, with all feeds passing $\chi^2$-tests at expected levels.

The PCA filtering (both in the time and map domain) constitutes the only non-linear processing in the pipeline. Non-linear filtering makes it more difficult to estimate the resulting signal bias and transfer function. In Sect.~\ref{sec:pca_linearity} we demonstrate that a PCA filter applied on a noisy matrix behaves linearly with respect to a very weak signal, and we find that any CO signal in the data is well within this safe regime.
\begin{figure}
    \centering
    \includegraphics[width=\linewidth]{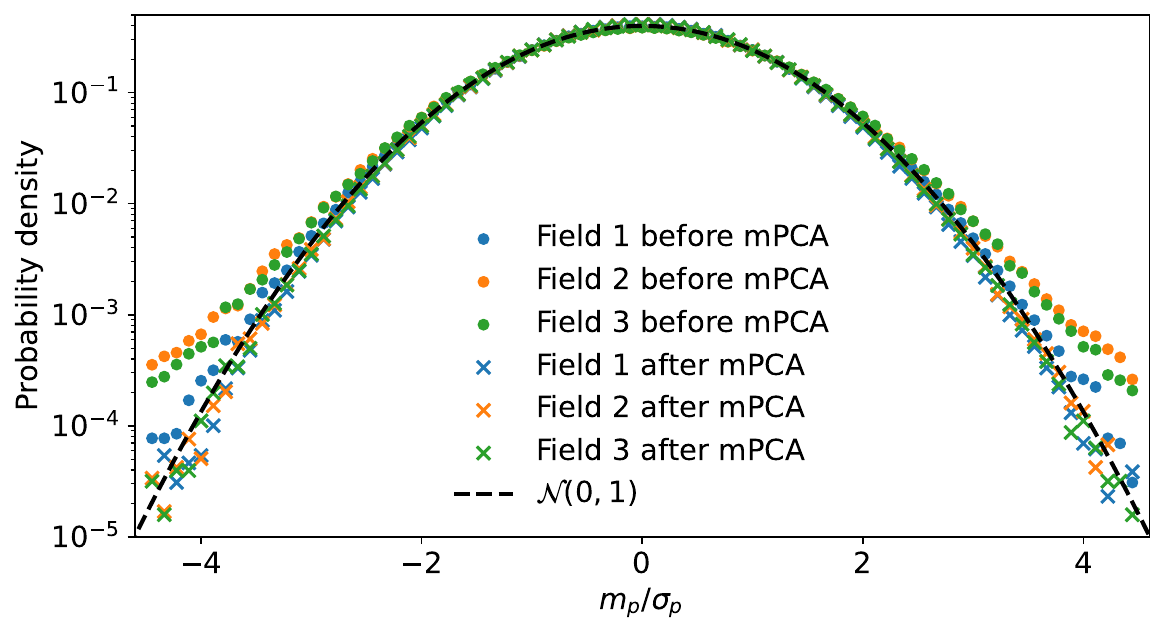}
    \vspace{-4mm}\caption{Histogram of all map pixel temperature values across all feeds and frequencies, divided by their white noise uncertainty. The three fields are shown separately, before and after application of the map PCA filter. A normal distribution is shown in black; a completely white noise map will trace this distribution. All three fields show excess high-significance pixels before the map PCA. After the filter, all three fields fall slightly below the normal distribution on the wings, because of the slight over-subtraction of noise at various stages in the pipeline.}
    \label{fig:map_pixels_hist}
\end{figure}

\subsection{Final maps}
Figure \ref{fig:map_pixels_hist} shows the distribution of map voxel values for all three fields, in units of significance, before and after the map-PCA. Before the map-PCA, the distribution shows a clear excess on the tails, while the cleaned maps are very consistent with white noise. This is expected and desired, as the CO signal is so weak that individual frequency maps are still very much dominated by the system temperature. The noise level in the maps is, actually, about 2.5\% lower than expected from the white noise uncertainty, due to the filtering in the pipeline. This effect can be seen in Fig.~\ref{fig:map_pixels_hist}, with the histograms falling slightly below the normal distribution.

\begin{figure}
    \centering
    \includegraphics[width=\linewidth]{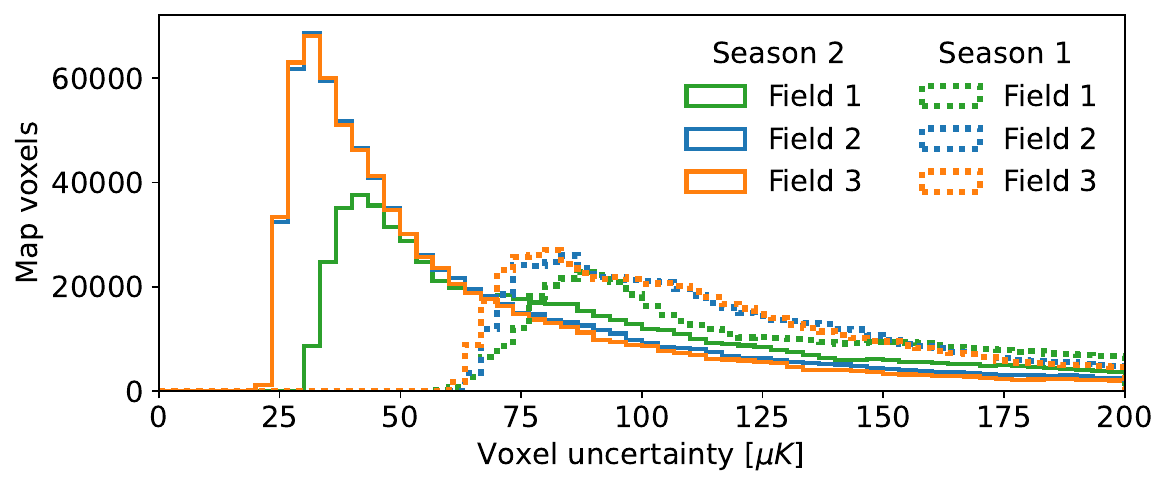}
    \vspace{-4mm}\caption{Histograms comparing the map voxel uncertainties of this work (Season 2), and the ES publications (Season 1). The voxel values are for feed-coadded maps.}
    \label{fig:map_rms_hist}
\end{figure}

Figure \ref{fig:map_rms_hist} shows the distribution of voxel uncertainties over the three fields for this work and our ES maps. Each voxel has an approximate size of $2\times\SI{2}{arcmin}$, that, together with the frequency direction, corresponds to a comoving cosmological volume of ${\sim}3.7\times 3.7\times \SI{4.1}{Mpc^3}$. For Fields 2 and 3, the high sensitivity $<\SI{50}{\mu K}$ region corresponds to a comoving cosmological cube of around $150\times 150\times \SI{1000}{Mpc^3}$ per field. Combining all three fields, Season 2 has one million voxels with an uncertainty $<\SI{50}{\mu K}$, compared to one million voxels below $<\SI{125}{\mu K}$ for Season 1. The footprint of the final maps have increased slightly in size because of the wider scan pattern of Season 2b.

The sensitivity increase per field over Season 1 is 2.0, 2.6, and 2.7, for Fields 1, 2, and 3, respectively. Fields 2 and 3 are now the highest sensitivity fields, while Field 1 is noticeably worse, from larger losses to data selection, especially Moon and Sun sidelobe pickup. The uncertainties are estimated from Eq.~(\ref{eqn:map_sigma}), and correspond well to the noise level observed in the map, as we saw from Fig.~\ref{fig:map_pixels_hist}. A figure showing the uncertainties across the fields on the sky can be found in Appendix \ref{app:maps}, together with a subset of the final maps.

\section{Data selection}
\label{sec:data_selection}
In addition to a three-fold increase in observational hours, the second season also features a similar increase in data retention compared to the ES results. Table \ref{tab:data_loss} compares the data loss in the ES results and this work. The table is split into three parts, namely 1) observational losses, 2) time- and map-domain losses, and 3) power spectrum domain losses. The Season 2 column only relates to data taken during Season 2, but we have also reprocessed Season 1 data with the new pipeline for the final results.

\subsection{Observational data retention}
The first three rows of Table \ref{tab:data_loss} are observational inefficiencies that have been corrected since the first season. $E_\mathrm{scan}$ constitutes the fraction of scans that were performed in constant elevation mode, as opposed to Lissajous scans that were cut due to large systematic effects (see Sect.~\ref{sec:Lissajous}). $E_\mathrm{feed}$ is the fraction of functioning feeds, and $E_\mathrm{el}$ is the fraction of data taken at elevation $35^\circ$--$65^\circ$ (see Sect.~\ref{sec:elevation}). Since Season 1, we no longer observe in Lissajous mode, all feeds are functional and the observing strategy has been optimized to maximize $E_\mathrm{el}$. As a result, the total data retention from these three cuts, which was 32\% in the first season, is now 100\%.

\subsection{Time and map domain data selection}
In the next section in Table~\ref{tab:data_loss}, $E_\mathrm{freq}$ refers to the frequency channel masking performed in \texttt{l2gen}, as discussed in Sect.~\ref{sec:l2gen:masking}. The masking algorithm itself is virtually identical to in ES, with a few changes. The shifting of aliased power into channels outside the nominal frequency range (Sect.~\ref{sec:aliasing}) means that, from Season 2b onward, we recover the $8\%$ of channels masked in Season 1 and 2a. The inclusion of the new per-feed PCA filter in \texttt{l2gen} results in slightly fewer data being masked by data-driven tests. However, we have also increased the number of manually flagged channels that seem to be performing sub-optimally, leaving us with a $E_\mathrm{freq}$ data retention only slightly higher than for ES.

Next, $E_\mathrm{stats}$ constitutes the cuts performed in the \texttt{accept\_mod} script, that discards scans based on different housekeeping data and summary statistics of the scans. There are over 50 such cuts in total, most of them removing a small number of outlier scans. Upon the completion of the Season 2 null test framework \citep{Stutzer_2024}, null tests failed on five scan-level parameters. Cuts on these parameters were tightened or implemented in \texttt{accept\_mod}, and the null tests now pass. The five new or tightened cuts are: 1) any rain during the scan; 2) wind speeds above $\SI{9}{m/s}$; 3) high average amplitude of the fitted TOD PCA components; and 4--5) outliers in the $f_\mathrm{knee}$ of the 0th and 1st order $1/f$ filter components\footnote{The $1/f$ filter fits the time-dependent components $c_t^0$ and $c_t^1$, primarily picking up correlated noise and changes in the atmosphere. We perform a $1/f$ fit to the components as functions of time, and cut when the $f_\mathrm{knee}$ falls outside the typical range of values}.
Additionally, all other \texttt{accept\_mod} cuts from ES are continued, and the surviving data fraction has therefore fallen noticeably, from 57.4\,\% to 33.6\,\%. No attempt has yet been made to tune these cuts, presenting us with future potential for increased data retention.

Finally, $E_{\chi^2_{P(k)}}$ is the last scan-level cut. Each scan is binned to a very low-resolution 3D map, and a series of $\chi^2$-tests are performed on different 2D and 3D auto power spectra calculated from these maps. In the Season~2 pipeline, this cut is removed entirely, for two reasons. Firstly, we found little evidence that it helped us pass null tests or remove dangerous systematic errors from the final data. Secondly, we found it difficult to calculate robust pipeline transfer functions for each individual power spectrum, as individual scans might vary a lot in sky footprint and pointing pattern. We therefore saw little reason to keep this cut in the pipeline.

\definecolor{verylightgray}{rgb}{0.9, 0.9, 0.9}
\newcolumntype{C}[1]{>{ \centering\arraybackslash}p{#1}}
\begin{table}[t]
    \setlength{\tabcolsep}{6pt} 
    \renewcommand{\arraystretch}{1.4} 
    \caption{\label{tab:data_loss} Data retention overview.}
    \centering
    \small
    \begin{tabular}{m{0.5cm} C{1.15cm} C{1.15cm} m{4.5cm}}
        \hline
        \hline
            & Season 1 & Season 2 & Explanation \\
            \hline
        $E_\mathrm{scan}$   & 50.0\% & 100.0\% & Retained scans (CESs) \\
        $E_\mathrm{feed}$   & 84.2\% & 100.0\% & Functional feeds \\
        $E_\mathrm{el}$     & 75.6\% & 100.0\% & Inside good elevation range \\
        \hline
        \rowcolor{verylightgray}
        $\vec{E_\mathrm{obs}}$    & $\mathbf{31.8\%}$ & $\mathbf{100.0\%}$ & Observational data retention\\
        \hline
        $E_\mathrm{freq}$   & 72.8\% & 74.3\% & Frequency masking in \texttt{l2gen}. \\
        $E_\mathrm{stats}$  & 57.4\% & 33.6\% & Cuts on accept-mod statistics\\
        $E_{\chi^2_{P(k)}}$ & 72.2\% & 100.0\% & Per-scan auto-PS $\chi^2$-test \\
        \hline
        \rowcolor{verylightgray}
        $\vec{E_\mathrm{cuts}}$    & $\mathbf{30.1\%}$ & $\mathbf{24.9\%}$ & Map-level data retention\\
        \hline
        $E_{\chi^2_{C(k)}}$ & 52.4\% & 100.0\%  & Cross-spectrum $\chi^2$-test \\
        $E_{C(k)}$          & 94.7\% & 75.0\% & Cross-spectrum auto combinations\\
        \hline
        \rowcolor{verylightgray}
        $\vec{E_\mathrm{PS}}$    & $\mathbf{49.6\%}$ & $\mathbf{75.0\%}$ & Retained data at PS-level \\
        \hline
        \hline
        \rowcolor{verylightgray}
        $\vec{S_\mathrm{tot}}$    & $\mathbf{6.8\%}$ & $\mathbf{21.6\%}$ & Final PS-domain sensitivity, calculated as $S_\mathrm{tot} = \sqrt{E_\mathrm{obs}^2 E_\mathrm{cuts}^2 E_\mathrm{PS}}$ \\
        \hline
    \end{tabular}
    \vspace{2mm}\tablefoot{Surviving fraction of data for different filtering steps of the pipeline. The left column shows the values used for the ES analysis, and the right column shows this work. The first 3 rows show individual data losses to observational constraints, that are combined in the gray row below. The three next rows show data retention to time and map domain cuts, again combined below. Finally, the next two rows show the losses in the power spectrum domain, also combined. The last row, $S_\mathrm{tot}$ shows the final fraction of theoretical power spectrum sensitivity from the combined data retention (see the text for details). The losses are multiplicative, such that multiplying $E$ for all the individual losses gives the retained data fractions shown in gray.}
\end{table}

\subsection{Power spectrum level data selection}
The last section of Table \ref{tab:data_loss} shows the fraction of data retained after cuts in the power spectrum domain; details on the power spectrum methodology are described by \citet{Stutzer_2024}. In summary, we calculate pseudo cross spectra between different groups of feeds and across pointing elevations and then average these spectra to get the CO power spectrum. However, some of the cross-spectra are discarded before averaging, and this is the loss discussed in this section. This loss in the power spectrum domain has to be tracked separately from data loss in the map and TOD domain, as the losses cannot naively be added together. Since map values are squared when calculating the power spectrum, so is the map-domain data volume when calculating the power spectrum sensitivity. The total power spectrum sensitivity is therefore calculated as $S_\mathrm{tot} = \sqrt{E_\mathrm{obs}^2E_\mathrm{cuts}^2 E_\mathrm{PS}}$.

In this table section, $E_{\chi^2_{C(k)}}$ constitutes a $\chi^2$-test on the individual feed-feed cross-spectra and cuts away any cross-spectrum with an average significance above $5\sigma$. In ES, we lost around half the data to this cut. The cut has now been entirely removed, for several reasons. First of all, we now have a much more rigorous null test framework, and find that we pass all null tests without these cuts. Secondly, we strongly prefer moving all data-inferred cuts to a point as early in the pipeline as possible, to reduce any potential biasing effects. It is therefore a considerable pipeline improvement over ES that we now perform no data-inferred cuts in the power spectrum domain.

Finally, $E_{C(k)}$ is the fraction of the cross power spectra that are not auto-combinations between the same feeds. In ES we performed cross-spectra between all 19 feeds, which resulted in a loss of $19$ out of $19\times19$ cross-spectra, or $5.3\%$. We now calculate cross-spectra between four groups of feeds, for better mitigation of systematic effects and improved overlap, resulting in a loss of 4 out of $4\times4$ feeds, or 25\,\%. This is a theoretical approximation of the sensitivity, as the varying degrees of overlap between different feeds will interplay with the sensitivity.

\subsection{Future prospects for data selection}
Combining the retained map-level data with the retained PS-level data we keep $21.6\%$ of the theoretical sensitivity, compared to $6.8\%$ for ES, a more than three-fold increase. Most of this increase comes from much higher observational data retention $E_\mathrm{obs}$, and the removal of the $\chi^2_{P(k)}$ and $\chi^2_{C(k)}$ cuts. We now have no data-driven cuts in the power spectrum domain, where the signal is the strongest, leaving us less susceptible to signal bias.

In order to pass null tests and allow for the removal of other cuts, the data retention after cuts on \texttt{accept\_mod} statistics, $E_\mathrm{stats}$, has decreased quite substantially. We erred on the side of caution when introducing the new cuts, and once the data passed all the null tests we made no attempt at reclaiming any data from these cuts. In future analysis, we are therefore confident that better tuning of these parameters, assisted by an even better understanding and filtering of systematic effects, will allow us to substantially increase the amount of data retained at this step.

The numbers in Table \ref{tab:data_loss} are averages across fields, feeds, and scans, and the combined data retentions, $E_\mathrm{map}$, $E_\mathrm{PS}$ and $S_\mathrm{tot}$ are for simplicity calculated by naively multiplying together the individual retentions. This ignores certain complications, such as correlations between the cuts, and the actual sensitivity might therefore differ slightly. It should also be noted that the right column constitutes the efficiency of Season 2 data in combination with the Season 2 pipeline, and re-analysis of Season 1 does not reach a $S_\mathrm{tot}$ of 21.6\%, as the losses to $E_\mathrm{obs}$ will still apply even with the improvements to the pipeline.

\section{Pipeline signal impact and updated transfer functions}\label{sec:TF}
The final maps are biased measurements of the CO signal, due to signal loss incurred in observation and data processing, leading to a biased power spectrum. This effect can be reversed by estimating a so-called transfer function $T(k_\parallel, k_\perp)$, that quantifies this signal loss at different scales. We separate the angular modes $k_\perp$ and the frequency (redshift) modes $k_\parallel$, as the impact on the CO signal is usually very different in these two dimensions. In this section, we present updated versions of the three relevant transfer functions:
\begin{itemize}
    \item The pipeline transfer function $T_\mathrm{p}(k_\parallel, k_\perp)$: The time- and map-domain processing will inevitably remove some CO signal.
    \item The beam transfer function $T_\mathrm{b}(k_\perp)$: The size and shape of the beam will suppress signal on smaller scales in the angular dimensions.
    \item The voxel window transfer function $T_\mathrm{v}(k_\parallel, k_\perp)$: The finite resolution of the voxels suppresses signal on both angular and redshift scales close to the size of the voxels.
\end{itemize}

\subsection{Updated beam and voxel window transfer functions}\label{subsec:beam_vox_window_tf}
\begin{figure}
    \centering
    \includegraphics[width=\linewidth]{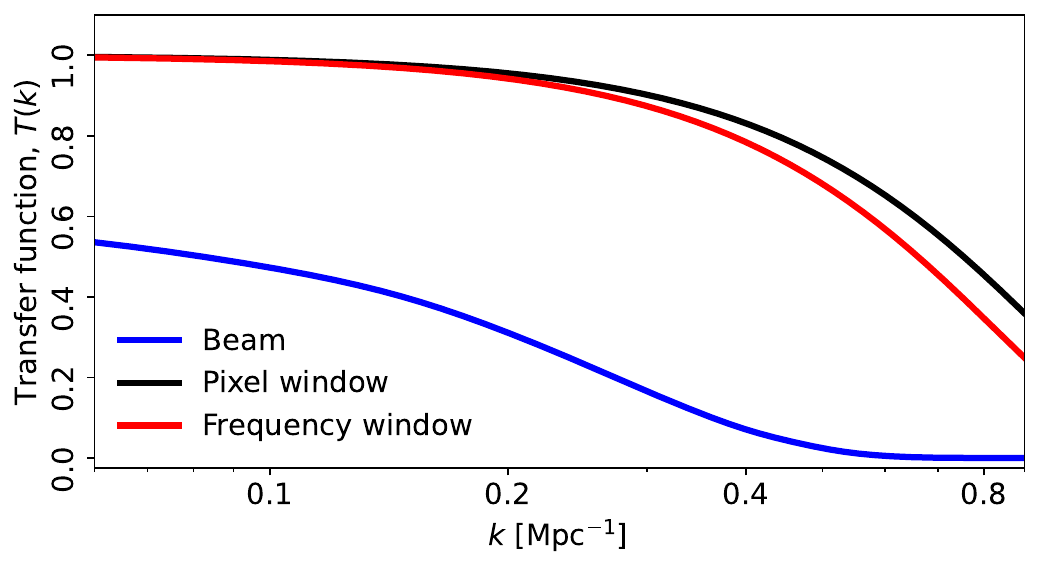}
    \vspace{-4mm}\caption{Effective one-dimensional transfer functions for the instrumental beam (blue curve), pixel window (black curve), and frequency window (red curve) resulting from spherical averaging over the corresponding two-dimensional transfer functions shown in Fig.~\ref{fig:2d_tf}.}
    \label{fig:1d_tf}
\end{figure}

\begin{figure*}
    \centering
    \includegraphics[width=\linewidth]{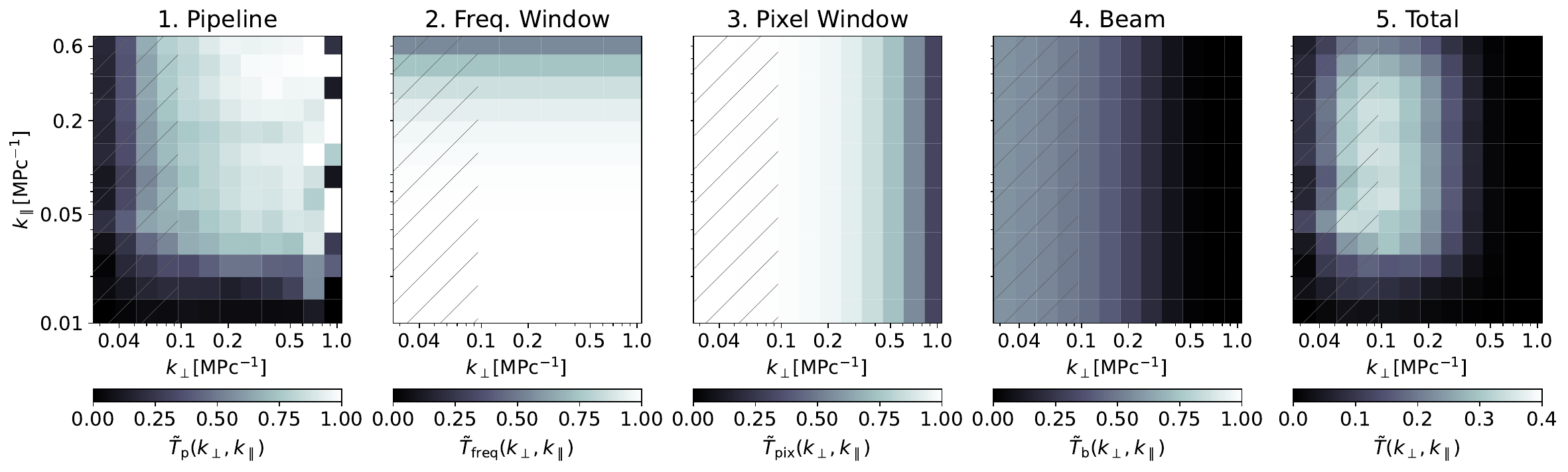}
    \vspace{-4mm}\caption{Effective transfer functions used in the COMAP pipeline. From left to right, the five panels show 1) the filter transfer function, $T_\mathrm{f}(\kvec)$, quantifying signal loss due to pipeline filters; 2) the pixel window transfer function $T_\mathrm{pix}(\kvec)$ resulting from binning the TOD into a pixel grid; 3) the frequency window transfer function, $T_\mathrm{freq}(\kvec)$, resulting from data down-sampling in frequency; 4) the beam smoothing transfer function, $T_\mathrm{b}(\kvec)$; and 5) the full combined transfer function, $T(\kvec)$, corresponding to the product of the four individual transfer functions. The striped region to the left is not used for our final analysis but is shown for completeness. We note that the leftmost panel has a colorbar that saturates at 0.4, unlike the other four.}
    \label{fig:2d_tf}
\end{figure*}

In the ES analysis, the beam and voxel window transfer functions were estimated using simulations \citep{Ihle_2022}. However, because the voxel grid and beam of the COMAP instrument are well understood we can also compute $T_\mathrm{b}(k_\perp)$ and $T_\mathrm{v}(k_\parallel, k_\perp)$ analytically. As the COMAP mapmaker simply uses nearest neighbor binning of the TOD into equispaced voxels the map is smoothed by a $\sinc^2(x)$ function along each map axis. Specifically, the voxel window can be expressed as\footnote{We note that we use the convention where $\sinc(x) = \frac{\sin(\pi x)}{\pi x}$.}
\begin{equation}
    T_\mathrm{v}(k_\parallel, k_\perp) = T_\mathrm{freq}(k_\parallel)T_\mathrm{pix}(k_\perp) = \sinc^2\qty(\frac{\Delta x_\parallel k_\parallel}{2\pi}) \sinc^2\qty(\frac{\Delta x_\perp k_\perp}{2\pi}),
\end{equation}
where $\Delta x_\perp$ and $\Delta x_\parallel$ are the voxel sizes in angular and frequency directions. Specifically, we have voxel resolutions of $\Delta x_\perp \approx 3.7\,\mathrm{Mpc}$ and $\Delta x_\parallel \approx 4.1\,\mathrm{Mpc}$. We note that since the angular pixel window is approximately radially symmetric we have approximated $T_\perp(k_\perp) \approx T_\mathrm{RA}(k_\mathrm{RA}) \approx T_\mathrm{Dec}(k_\mathrm{Dec})$. Both the perpendicular and parallel voxel transfer functions can be seen in Figs.~\ref{fig:1d_tf} and \ref{fig:2d_tf}.

In principle, we could reduce the voxel window signal impact on smaller scales in both the angular and frequency dimensions by binning the maps into higher resolution voxels, shifting the decline of $T_\mathrm{freq}(k_\parallel)$ and $T_\mathrm{pix}(k_\perp)$ to higher $k$-values. In practice, however, the angular voxel window applies at a scale where the beam transfer function already suppresses the signal beyond recovery. Similarly, line broadening is expected to heavily attenuate the CO signal above $\SI{{\sim}1}{Mpc^{-1}}$ \citep{Chung_2024}, although the exact extent of line-broadening depends on galaxy properties that are not yet well constrained. Additionally, it would be more computationally costly to perform the analysis in higher resolution.

Next, given the radial beam profile $B(r)$ \citep[see Fig. 2 of][]{Ihle_2022} and the convolution theorem we can obtain the beam transfer function as
\begin{equation}
    T_b(k_\perp) = |\mathcal{F}\{B(r)\}|^2,
\end{equation}
where $r$ is the radius from the beam center, and $\mathcal{F}$ is the (2D) Fourier transform. As we assume the telescope beam to be radially symmetric the resulting beam smoothing transfer function will be a function of just $k_\perp = \sqrt{k_\mathrm{RA}^2 + k_\mathrm{Dec}^2}$ giving $T_\mathrm{b}(k_\perp)$. The main-beam efficiency is taken into account in the same manner as \cite{Ihle_2022} prior to computing the Fourier transform of the beam. As we can see in Figs.~\ref{fig:1d_tf} and \ref{fig:2d_tf} the beam is by far the most dominant effect limiting our ability to recover the signal at smaller scales.

\subsection{The signal injection pipeline}
\label{sec:signal_injection_pipeline}
The last transfer function is that of the pipeline, that will inevitably remove some CO signal from the data. We estimate this impact by a signal injection pipeline, similar to what was done in ES \citep{Foss_2022}.

We inject a simulated CO signal into the real Level 1 data before any filtering. The data are then propagated through the entire pipeline as usual, and the resulting mock observations are then compared to the known, unfiltered input signal to estimate the pipeline transfer function. We chose to inject the simulations into the actual data, instead of simulating the entire observation, in order to mimic the real systematic error and noise properties as closely as possible.

For the simulations we use approximate cosmological dark-matter-only simulations using the peak-patch method of \cite{bond:1996} with updates by \cite{stein:2019}. These are subsequently populated with CO emission using the COMAP fiducial model derived in~\cite{comap-V} (`UM+COLDz+COPSS'), that describes CO luminosities as a function of dark matter halo masses, $L_\mathrm{CO}(M_\mathrm{halo})$. These simulated mock maps are then boosted by a factor of 20, to recover a less noisy transfer function. This is counter-weighted by splitting the full dataset in 10 subsets, and passing each of these separately though the pipeline. We denote the maps as $\vec{s}^\mathrm{mock}_{\nu, \theta}$, where $\nu$ and $\theta$ are the frequency (redshift) and angular (pixel) dimensions, respectively. The TOD pipeline is quantitatively unaffected by the injection of this weak CO signal -- even after a factor 20 boost, the brightest CO pixel in the simulation is still more than four orders of magnitude below the system temperature. In the map-domain, the S/N is much higher, and the implications of injecting a boosted signal are discussed in Sect. \ref{sec:pca_linearity}, where we conclude that the map-PCA filter also behaves predictably for our chosen boost strength.

Using the real telescope pointing $\tens{P}$, and estimated gain $\tens{G}$ and beam $\tens{B}$, we get the signal-injected Level 1 data as
\begin{equation}
    d^\mathrm{mock}_{t, \nu} = \tens{G}\tens{P}\tens{B} s^\mathrm{mock}_{\theta, \nu} + n_{t, \nu}, 
\end{equation}
where $n_{\nu, t}$ represents the actual Level~1 data, that acts as the noise term with respect to the injected mock CO signal. In order to mimic the observed CO signal as closely as possible, we also beam-smooth the maps used for the signal injection. The mock data $d^\mathrm{mock}_{\nu, t}$ are then filtered by the pipeline to produce a mock map:
\begin{equation}
\begin{aligned}
    m^\mathrm{mock}_{\theta, \nu} &= f_\mathrm{map}\qty[M\qty(f_\mathrm{TOD}\qty[d^\mathrm{mock}_{\nu, t}])]\\ &= f_\mathrm{map}\qty[M\qty(f_\mathrm{TOD}\qty[\tens{G}\tens{P}\tens{B}s^\mathrm{mock}_{\nu, t} + n_{\nu, t}])],
\end{aligned}
\end{equation}
where we let $f_\mathrm{TOD}$ represent all time-domain filtering, $M$ represents a noise-weighted binned map-maker, as described in Sect.~\ref{sec:mapmaker}, and $f_\mathrm{map}$ represents the map-domain PCA filter. Examples of resulting maps are shown in Appendix \ref{app:signal-injection}.

To make sure that the reference CO simulation $\vec{s}^\mathrm{mock}_{\nu, \theta}$ is directly comparable to the filtered maps, we also perform some of the same treatment on it: we beam-smooth it, read it into a TOD with the real telescope pointing, and bin the TOD back into a map with the same resolution as the real maps. The difference, however, is that this is done completely without noise, and we do not apply any of the filters. We can write this as
\begin{equation}
    \hat{s}^\mathrm{mock}_{\nu, \theta} = M\qty(\tens{G}\tens{P}\tens{B}s^\mathrm{mock}_{\nu, \theta}).
\end{equation}
Doing it this way means that we isolate the filter transfer function, and the effect of the beam and pixelation are not included. This is intentional, as we already estimated these impacts analytically in Sect. \ref{subsec:beam_vox_window_tf}.

From these maps, we can now write the filter transfer function as 
\begin{equation}
    \label{eq:filter_tf}
    T = \frac{C(\vec{m}^\mathrm{mock}, \hat{\vec{s}}^\mathrm{mock})}{P(\hat{\vec{s}}^\mathrm{mock})},
\end{equation}
where the cross-spectrum, $C$ in the numerator between the filtered mock data, $\vec{m}^\mathrm{mock}$, and the unfiltered mock signal, $\hat{\vec{s}}^\mathrm{mock}$, picks up all common signal modes after filtering while canceling residual systematic effects and noise in the mock data. The cross-spectrum is divided by the unfiltered signal auto spectrum, $P_\kvec(\hat{\vec{s}}^\mathrm{mock})$, to obtain a filter transfer function $T$.

Equation~(\ref{eq:filter_tf}) represents a more robust estimator of the transfer function than the one employed in ES (\cite{Foss_2022} Eq. 34), both because it does not require an accurate estimation of the noise power spectrum, and because using the cross-spectrum estimator as opposed to the auto-spectrum estimator makes it less susceptible to picking up signal in the data that does not originate from the injected CO (e.g. from systematic effects).

For the signal injection, we use all scans from Seasons 1 and 2a for Field 2. Preliminary analysis of the transfer functions of Fields 1 and 3, and the slower pointing scans of Season 2b, show that they are very similar, especially in the $k$-regime included in this work. As mentioned in the beginning of the section, we divide the scans into 10 random and equally large parts. Different dark matter halo simulations are injected into each part. This both reduces the impact of sample variance in the simulations and allows us to boost the signal a bit more without having to worry about PCA non-linearity (see the next section). We then average over the 10 resulting transfer functions, to get the final transfer function estimate.

\subsection{Updated filter transfer function}
The left-most panel of Fig.~\ref{fig:2d_tf} shows the full COMAP pipeline transfer function, as described in the previous section, in parallel and perpendicular directions (i.e., redshift and angular scale, respectively). The Season 2 publications \citep{Stutzer_2024, chung:2024} exclude some of the larger angular scales accessible in the maps due to concerns about mode mixing and unconstrained modes due to poor overlap. For reference, the cutoff value for $k_\perp$ at $\SI{0.93}{MPc^{-1}}$ corresponds to angular scales of $\SI{36.4}{arcmin}$.

\begin{figure*}
    \centering
    \includegraphics[width=\linewidth]{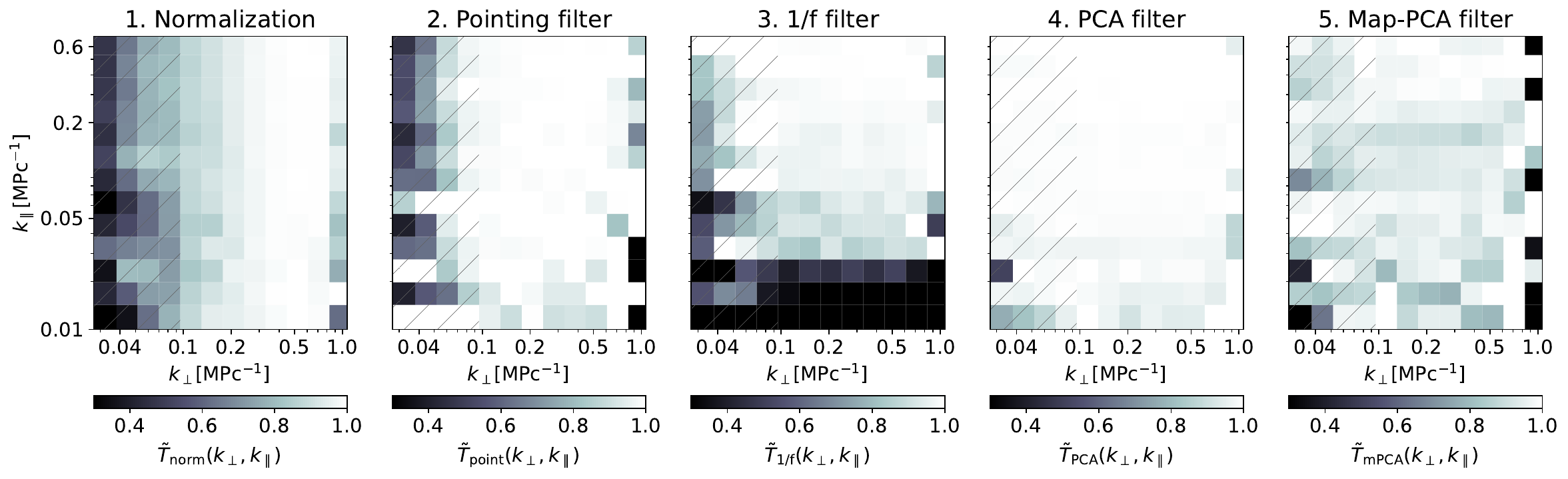}
    \vspace{-4mm}\caption{Transfer functions for each of the five individual filters used in the pipeline.  
      The normalization and pointing filters suppress large angular scales, while the $1/f$ filter almost entirely eliminates parallel modes larger than $k=\SI{0.02}{Mpc^{-1}}$. The TOD PCA filter has almost no impact on the signal, primarily because so few modes are subtracted. The map PCA has a more noticeable signal loss, but it remains relatively scale-independent because the signal is still weak enough (see Sect.~\ref{sec:pca_linearity}). The total pipeline transfer function shown in the leftmost panel of Fig.~\ref{fig:2d_tf} is the product of these. The striped region to the left is not used in Season 2 results.}
    \label{fig:TF_perfilter}
\end{figure*}

Figure \ref{fig:TF_perfilter} shows the individual contributions of each filter to the full transfer function. We are somewhat limited on large angular scales by the normalization and pointing filters, and very limited on large redshift scales by the $1/f$ filter. The right-most column is noisy because the beam suppresses most of the signal at these small scales.

We also note that a small issue was discovered with the transfer function analysis published in \cite{Ihle_2022}, relating to how the mock signal was interpolated when injected into the TOD. The effect was that the transfer functions from ES was slightly underestimated, and the transfer function from \cite{Ihle_2022} peaked at around 0.8. This issue has now been solved, and the new transfer function peaks correctly at almost 1.0.

\subsection{Linearity of PCA filtering}
\label{sec:pca_linearity}
All filters except the various PCAs constitute linear operations on the data. Linearity makes transfer function estimation much simpler, as neither the choice of CO signal model nor its level with respect to the noise impacts the resulting transfer function. For the PCA filters, both these factors could in principle impact the shape of the transfer function.

To quantify the sensitivity of the transfer function to such factors, we constructed a simplified version of the signal-injection pipeline. In order to be able to run many simulations, we made the following alterations to the pipeline:
\begin{itemize}
    \item We bypass the time domain, and perform the signal injection in the map domain. This is the domain where the CO S/N is the strongest, and the map-domain processing is much more computationally efficient than that for the time-domain. The map mocks are the same as in Sect.~\ref{sec:signal_injection_pipeline}, with some boost factor $b$.
    \item Instead of the real map, we use white noise simulations, drawn from the white noise uncertainty of Field 2.
    \item The resulting transfer function is  calculated using Eq.~(\ref{eq:filter_tf}), and then averaged across the simulations and feeds.
\end{itemize}

\begin{figure}
    \includegraphics[width = \linewidth]{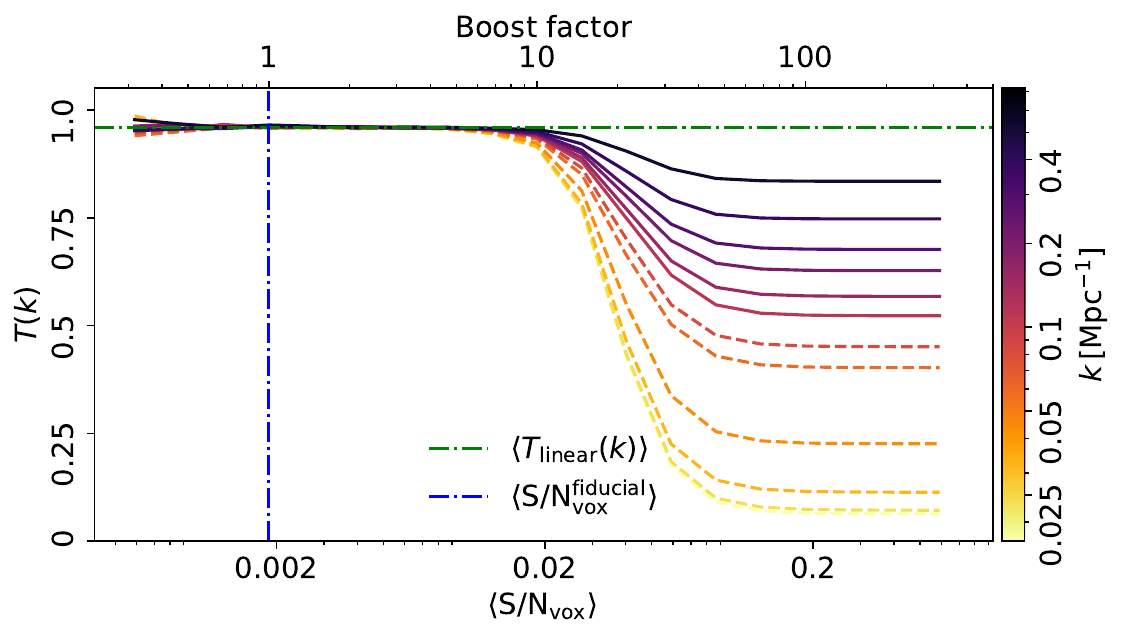}
    \vspace{-4mm}\caption{Map PCA transfer function $T(k)$ as a function of the voxel S/N, with the different $k$-scales shown as differently colored lines. The equivalent boost used to the fiducial CO model is shown as the top $x$-axis. Modes with $k$-values that fall within the analysis bounds are shown as solid, while modes outside the scope of COMAP Season 2 are shown as striped. The horizontal green line shows the average transfer function in the low-S/N regime, which is 0.96. The vertical blue line shows the S/N that the maps would have if the fiducial model of \cite{comap-V} perfectly described the true CO signal.}
    \label{fig:mPCA_TF_linearity}
\end{figure}

This process is repeated for 40,000 noise realizations for each of the 19 feeds, with boosts between 0.3 and 300 relative to the fiducial CO model of \cite{comap-V}. The variation of the resulting average transfer functions with boost strength can be seen in Fig.~\ref{fig:mPCA_TF_linearity}. The figure shows two distinct regimes. In the low S/N regime to around S/N=0.02 (boost 10), the PCA filter behaves linearly with respect to the CO signal. This is demonstrated by the independence of the transfer function to the S/N in this regime. Additionally, all the $k$-points lie on top of each other around a value of 0.96, meaning that all scales are suppressed at the same level because the PCA is simply fitting and subtracting random white noise. In the second regime, at high S/N, the transfer function is strongly scale-dependent but flattens out as the signal dominates the noise.

In conclusion, for a noisy matrix with an accompanying signal, a PCA filter behaves linearly with respect to the signal for a sufficiently weak signal. For us this means that when estimating the transfer functions, we need to use a sufficiently low boost value to avoid biasing our estimate of the transfer function. The individual data chunks used to estimate the transfer function are well within this linear regime, at an S/N of 0.004. The actual CO signal is of course of unknown amplitude, but assuming the fiducial model of \cite{comap-V} results in an S/N of 0.002. If the CO signal were even close to the unsafe regime of S/N~$> 0.02$, we would already have made a strong detection of it, as this would correspond to a 100 times brighter power spectrum than the fiducial model. We also note that in the future, as the experiment's sensitivity increases, we can simply perform the map PCA on sub-divisions of the data to keep the S/N low, as we already do on Season 1 + 2a and Season 2b, due to their differing pointing strategy.

This analysis was performed on the map-domain PCA filter, as the map domain is where the CO S/N is the strongest. Equivalent analysis has been performed for the TOD PCA filter, but the CO signal is so weak at the per-scan level that a boost factor of 2000 or greater is required to make it behave non-linearly with respect to the CO signal. The final transfer function is currently estimated jointly for all filters, but in future work we intend to estimate the transfer function for the mPCA separately from the other filters, gaining higher sensitivity on the linear parts of the pipeline that do not require as low a boost.

\section{Summary and conclusions}
\label{sec:conclusions}
We have presented the improvements in data analysis, filtering, and data selection that have enabled us to increase the power spectrum sensitivity to $21.6\%$ of the theoretical maximum, up from $6.8\%$ in the ES publications. Combined with an increased integration time, the two most sensitive Season~2 fields both have a voxel uncertainty of $<\SI{50}{\mu K}$ across ${\sim}1.5^\circ \times 1.5^\circ$ patches on the sky. Across the three fields, this corresponds to a 2.5 times decrease in the total map uncertainty.

The largest increase in data retention comes from improvements in observational strategy. We now solely observe using CES, whereas in Season 1 50\% of observations used Lissajous scans that proved prone to systematic effects and were not included in our ES analysis. Additionally, we now observe within elevation boundaries of $35^\circ$--$65^\circ$, a region with minimal gradients in ground sidelobe pickup. In Season 1, $25\%$ of scans fell outside this range and were discarded. 

Additional increases in data retention have come from the removal of data cuts. In Season 1, power spectrum $\chi^2$ tests were performed both on individual scan maps and on individual feed-feed cross-spectra before averaging them. These cuts removed (respectively) $28\%$ of scans and $48\%$ of cross-spectra, and are now no longer applied. This also reduces the possibility of signal bias owing to data-inferred cuts late in the pipeline.

In order to pass null tests and allow for the removal of other cuts, scan-level data cuts were implemented on six new housekeeping statistics, increasing the data lost to such cuts from $43\%$ to $66\%$. However, no attempt was made to reclaim data after the null tests had passed, and the necessity of the cuts carried over from the ES pipeline is largely untested. We are confident this number can be greatly reduced in future work.

The removal of data cuts was also made possible by better mitigation of systematic errors. The time-domain pipeline has numerous smaller improvements, such as a new per-feed PCA filter, dynamically determined number of PCA components, better masking of the $T_\mathrm{sys}$ spikes, and more manual masking of consistently problematic channels. Most impactful, however, was the introduction of a map-level PCA filter, that proved essential to dealing with a couple of pointing-correlated systematic errors that emerged due to increased sensitivity. We show that the map PCA filter suppresses these effects to below the noise level, and decreases the standard deviation of slightly smoothed maps by 67\%, to a level consistent with the expected white noise.

Although the PCA filters constitute non-linear filtering, we have shown that the PCA filters behave linearly with respect to any sufficiently weak signal. We find that the expected CO signal falls well within this regime, substantially simplifying transfer function estimation. We repeat the signal-injection pipeline transfer function estimation of the ES publications and ensure that the injected signal is also weak enough to maintain the PCA filters in their linear range. We have also replaced the simulation-based estimates of the beam and voxel windows transfer functions with more robust analytic expressions, improving their reliability at small scales.

COMAP has thus greatly increased its integration speed both through observational improvements, better processing, and reduced data cuts. Our final sensitivity retention of $21.6\%$ of the theoretical maximum still leaves significant room for improvement, and we aim to increase this further as the Pathfinder continues to observe.

\begin{acknowledgements}    
    We acknowledge support from the Research Council of Norway through grants 251328 and 274990, and from the European Research Council (ERC) under the Horizon 2020 Research and Innovation Program (Grant agreement No. 772253 bits2cosmology and 819478 Cosmoglobe)
    
    This material is based upon work supported by the National Science Foundation under Grant Nos.\ 1517108, 1517288, 1517598, 1518282, 1910999, and 2206834, as well as by the Keck Institute for Space Studies under ``The First Billion Years: A Technical Development Program for Spectral Line Observations''.

Parts of the work were carried out at the Jet Propulsion Laboratory, California Institute of Technology, under a contract with the National Aeronautics and Space Administration.

    HP acknowledges support from the Swiss National Science Foundation via Ambizione Grant PZ00P2\_179934. 

SEH and CD acknowledge funding from an STFC Consolidated Grant (ST/P000649/1) and a UKSA grant (ST/Y005945/1) funding LiteBIRD foreground activities.

This work was supported by the National Research Foundation of Korea(NRF) grant funded by the Korea government(MSIT) (RS-2024-00340759).

PCB was supported by the James Arthur Postdoctoral Fellowship.

DTC was supported by a CITA/Dunlap Institute postdoctoral fellowship for much of this work. The Dunlap Institute is funded through an endowment established by the David Dunlap family and the University of Toronto. Research in Canada is supported by NSERC and CIFAR.
    
    JGSL and NOS thank Sigurd~K.~N{\ae}ss for all the fruitful discussions in the office, and while biking through nature, during the last three years.

    This work was first presented at the Line Intensity Mapping 2024 conference held in Urbana, Illinois; we thank Joaquin Vieira and the other organizers for their hospitality and the participants for useful discussions.

    JGSL and NOS thanks the 2023 Stockholm Beam Mode workshop for discussions on beam convolution.

    \newline\newline \textit{Software acknowledgments}. Matplotlib for plotting  \citep{matplotlib:2007}; NumPy \citep{numpy:2020} and SciPy \citep{scipy:2020} for efficient numerics and array handling in Python; Astropy a community made core Python package for astronomy \citep{astropy:2013, astropy:2018, astropy:2022}; Multi-node parallelization with MPI for Python \citep{mpi4py:2005, mpi4py:2008, mpi4py:2011, mpi4py:2021}; Pixell (\url{https://github.com/simonsobs/pixell}) for handling sky maps in rectangular pixelization.
\end{acknowledgements}

\bibliographystyle{aa}
\bibliography{references}

\appendix

\section{Dynamic PCA threshold} \label{app:pca_threshold}
It is known that the largest singular value of a Gaussian $P \times N$ matrix with variance $\sigma^2$ can be approximated as ${\lambda \approx C(P,N) \cdot \sigma(\sqrt{P} + \sqrt{N})}$ for large matrices \citep{geman1980limit, rudelson2010non, vershynin2010introduction}, where $C(P,N)\approx 1$ is some correction-factor for which no reliable theoretical model exists. We, therefore, simulated 50,000 noise matrices and empirically solved for $C(P,N)$ within the relevant regime of $N$ from 5,000 to 30,000, and $P$ from 100 to 80,000, that captures all matrix sizes our pipeline will encounter. We find that the correction factor can, in our size range, be well-modeled as
\begin{align}
    C(P,N) = 1.00476 &- 0.00396 \cdot \log(P)\cdot \log(N) \notag\\
    &+ 0.0000876 \cdot \log(P)^2 \cdot \log(N)^2.
\end{align}
We have no doubt that this factor is unlikely to extrapolate sensibly beyond the range we tested it in, but that is of no concern. The relative error to the mean of our 50,000 simulations is less than 0.1\% within our defined bounds.

\section{The map-domain PCA filter}
\label{app:mpca_solution}
\subsection{The effect of noise-weighting on PCA}
This section expands on the PCA discussion of Sect. \ref{sec:mPCA}, and we keep the same variable and index names, for easier comparison. The first principal component $\vec{w}^1$ and its amplitude $\vec{a}^1$ of a PCA\footnote{In the regular PCA formalism, the eigenvectors $\vec{w}$ are typically unit vectors of length 1, which is not automatically the case throughout this section. However, $\vec{w}$ can be normalized to $1$ at any point by inversely adjusting $\vec{a}$. In a regular PCA, a (scalar) singular value is present in the solution. In the formalism presented here, there is no explicit singular value, and it can be absorbed into the amplitudes $\vec{a}$.} for a data-matrix $\vec{m}$ are the vectors that minimize 
\begin{equation}\label{eqn:mPCA4}
    \quad f(\vec{a}, \vec{v}) = \sum_\nu\sum_p (m_{\nu,p} - a_\nu w_p)^2.
\end{equation}
We might, however, want to find these vectors while weighting the elements of $\vec{m}$, for example, if it is non-uniformly noisy. If the weights themselves can be separated into an outer product, such that we have row and column uncertainties $\vec{\sigma}^\mathrm{row}$ and $\vec{\sigma}^\mathrm{col}$, this can trivially be done by inverse-variance weighting the matrix elements of the expression above. We now minimize 
\begin{equation}\label{eqn:mPCA2}
     f(\vec{a}, \vec{w}) = \sum_\nu\sum_p \frac{(m_{\nu,p} - a_\nu w_p)^2}{\sigma^\mathrm{row}_\nu \sigma^\mathrm{col}_p},
\end{equation}
which can be expanded to
\begin{equation}
    f(\vec{a},\vec{w}) = \sum_\nu\sum_p \qty(\frac{m_{\nu,p}}{\sqrt{\sigma^\mathrm{row}_\nu \sigma^\mathrm{col}_p}} - \frac{a_\nu}{\sqrt{\vphantom{\Big |}\sigma^\mathrm{row}_\nu}} \frac{v_p}{\sqrt{\sigma^\mathrm{col}_p}})^2.
\end{equation}
This is still a valid PCA problem, on the same form as Eq.~(\ref{eqn:mPCA4}), with $\vec{a}' = \vec{a}/\vec{\sigma}^\mathrm{row}$ and $\vec{w}' = \vec{w}/\vec{\sigma}^\mathrm{col}$ now being the amplitude and component we fit for. If $\vec{m}$ contains a feature that can be decomposed into an outer product, this will be recovered by $\vec{w} = \vec{w}'\vec{\sigma}^\mathrm{row}$ and $\vec{a} = \vec{a}'\vec{\sigma}^\mathrm{col}$.

However, if we want to weight every element of $\vec{m}$ with arbitrary uncertainty $\sigma_{\nu,p}$, this no longer holds. We can still write the problem simply as
\begin{equation}\label{eqn:mPCA3}
    f(\vec{a}, \vec{w}) = \sum_\nu\sum_p \frac{(m_{\nu,p} - a_\nu w_p)^2}{\sigma_{\nu,p}^2},\\
\end{equation}
   which can be expanded to
\begin{equation}
    f(\vec{a},\vec{w}) = \sum_\nu\sum_p \qty(\frac{m_{\nu,p}}{\sigma_{\nu,p}} - \frac{a_\nu}{\sqrt{\sigma_{\nu,p}}} \frac{w_p}{\sqrt{\sigma_{\nu,p}}})^2,
\end{equation}
but because $a_\nu/\sigma_{\nu,p}$ and $w_p/\sigma_{\nu,p}$ are now matrices and not vectors, this minimization problem is no longer a PCA. We therefore have no way of recovering the desired $\vec{a}$ and $\vec{w}$ corresponding to the matrix $\vec{m}$ if we perform the regular PCA algorithm on the matrix $m_{\nu,p}/\sigma_{\nu,p}$, for a general $\sigma_{\nu,p}$. We must therefore find a different way of minimizing \ref{eqn:mPCA3}, which is discussed in the following section.

\subsection{Generalization of the PCA algorithm}

With the generalization of Eq.~(\ref{eqn:mPCA3}), we can no longer utilize the usual methods of solving a PCA problem, such as the SVD. Instead we employ the technique suggested by \cite{tamuz2005correcting} and \cite{gabriel1979lower}, where we iteratively make improved guesses at $\vec{w}$ and $\vec{a}$:
\begin{enumerate}
    \item Make an initial guess at $\vec{a}$ and $\vec{w}$, either completely random or informed by some knowledge of the data.
    \item Solve for the optimal $\vec{a}$ while holding the current $\vec{w}$ constant by differentiating Eq.~(\ref{eqn:mPCA3}), holding $\dv{f(\vec{a},\vec{w})}{\vec{a}} = 0$, and solving for $\vec{a}$ as
    \begin{equation}
        a_\nu = \frac{\sum_p \frac{m_{\nu,p}w_p}{\sigma_{\nu,p}^2}}{\sum_p\frac{w_p^2}{\sigma_{\nu,p}^2}}.
    \end{equation}
    \item Given the new $\vec{a}$, calculate $\dv{f(\vec{a},\vec{w})}{\vec{w}} = 0$, and solve for the new optimal $w_p$ as
    \begin{equation}
        w_p = \frac{\sum_\nu \frac{m_{\nu,p}a_\nu}{\sigma_{\nu,p}^2}}{\sum_\nu\frac{a_\nu^2}{\sigma_{\nu,p}^2}}.
    \end{equation}
    \item Repeat 2. and 3. until the incremental changes in $\vec{a}$ and $\vec{w}$ are below some chosen threshold $\epsilon$.
\end{enumerate}
Although we cannot prove that this is a convex problem and that the optimal solution is guaranteed, we have never seen it converge to an unreasonable solution. Additional robustness can be achieved by repeating the fit with different initial guesses, and confirming that they converge to the same solution. The algorithm will converge on the same solution as the regular PCA in the case of uniform weights $\sigma_{\nu,p} = 1$, or where the weights can be perfectly decomposed into an outer product of weights in rows and columns, as in Eq.~(\ref{eqn:mPCA2}).

\subsubsection{Multiple components}
Using this decomposition, we can fit multiple components, similar to the ordered set of principal components in the PCA, by simply subtracting the previous components from the data. We then simply define $a_{i}^{(1)}$ and $m_{j}^{(1)}$ as the results from the previous section, and let
\begin{equation}
    m_{\nu,p}^{(1)} = m_{\nu,p} - a_\nu^1 w_p^1.
\end{equation}
We can then find $a_\nu^{(2)}$ and $w_p^{(2)}$ by performing the procedure from the previous section on $m_{\nu,p}^{(1)}$. This is again, in the case of $\sigma_{\nu,p} = 1$, entirely equivalent to finding the largest principal components of $m_{\nu,p}$.

\section{Start-of-scan and turn-around effects}\label{app:new-systematics}
\begin{figure*}
    \centering
    \includegraphics[width=\textwidth]{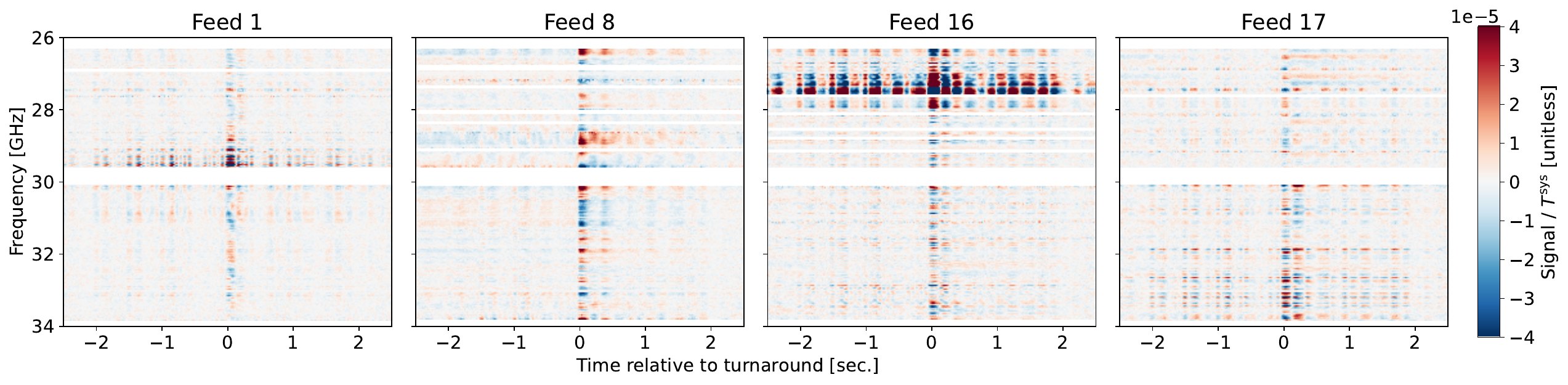}
    \vspace{-4mm}\caption{Filtered time-domain data stacked on the turn-around of the telescope, to emphasize the turn-around systematic effect, for four selected feeds. The data is an average across all turn-arounds of all available Level 2 scans. The data is divided by the system temperature of the data in each channel, and the values therefore represent the signal strength of the effect relative to the average white noise level in the scans.}
    \label{fig:turnaround-TOD}
\end{figure*}
As outlined in Sect. \ref{sec:new-systematics}, our increased sensitivity has revealed two systematic effects in our maps that were not discovered in our ES publications. We here explore these effects in more detail, especially in the time-domain, where these effects are easier to understand than in the map-domain. We note that the data shown in this section has not been filtered by the map-PCA filter, and the systematic effects demonstrated are (to the best of our analysis) not present in our final maps.

\subsection{The turn-around effect}

The turn-around effect is a sharp feature located around the azimuth edges of the scanning pattern, where the telescope turns. In Fields 2 and 3, which rise and set almost vertically across the sky (see Figure \ref{fig:ground_pickup}) and are therefore observed at almost the same angle at all times, this effect manifests as sharp edges on the top and bottom (i.e., the highest and lowest declinations) of the equatorial coordinate maps. This can be seen in the first and third rows of Fig. \ref{fig:mPCA_figures}.

To better understand this systematic effect, we have extracted the data around the telescope turn-arounds for all our scans, and stacked the result on the turn-around time. Figure \ref{fig:turnaround-TOD} shows the result for four selected feeds. For all four feeds, we see a feature that peaks around the turn but is also present both leading up to and after the turn. In the frequency direction the feature has a slow wave-like feature. The feature manifests differently in different feeds and frequencies but has in common that it is wave-like both in frequency and time and peaks in power around the turn-around. As the telescope turn-arounds are the regions with the highest acceleration, a likely origin of this effect is some standing wave induced by the mechanical vibrations of the azimuth drive. Some feeds show significantly stronger manifestations of the turn-around effect than others, but all feeds are affected to some extent, and no explanation has yet been found as to why feeds are affected differently.

Attempts have been made to model this effect in the time-domain. This has proven difficult, among other reasons because the effect is actually very weak compared to the noise level in a single scan. As seen from Fig.~\ref{fig:turnaround-TOD}, the effect peaks at more than four orders of magnitude below the noise temperature of the telescope. The effect is only visible in the final maps because it seems strongly coherent across different scans. However, because we have to combine thousands of scans in order to observe the effect, it is also difficult to assess if the effect is indeed perfectly coherent across all scans, or if we are simply observing the average impact of this effect. The efforts of modeling the effect in the time domain was also made somewhat moot by how effective the map-PCA was at removing the effect in the map-domain. 

\subsection{The start-of-scan effect}
\begin{figure}
    \centering
    \includegraphics[width=\linewidth]{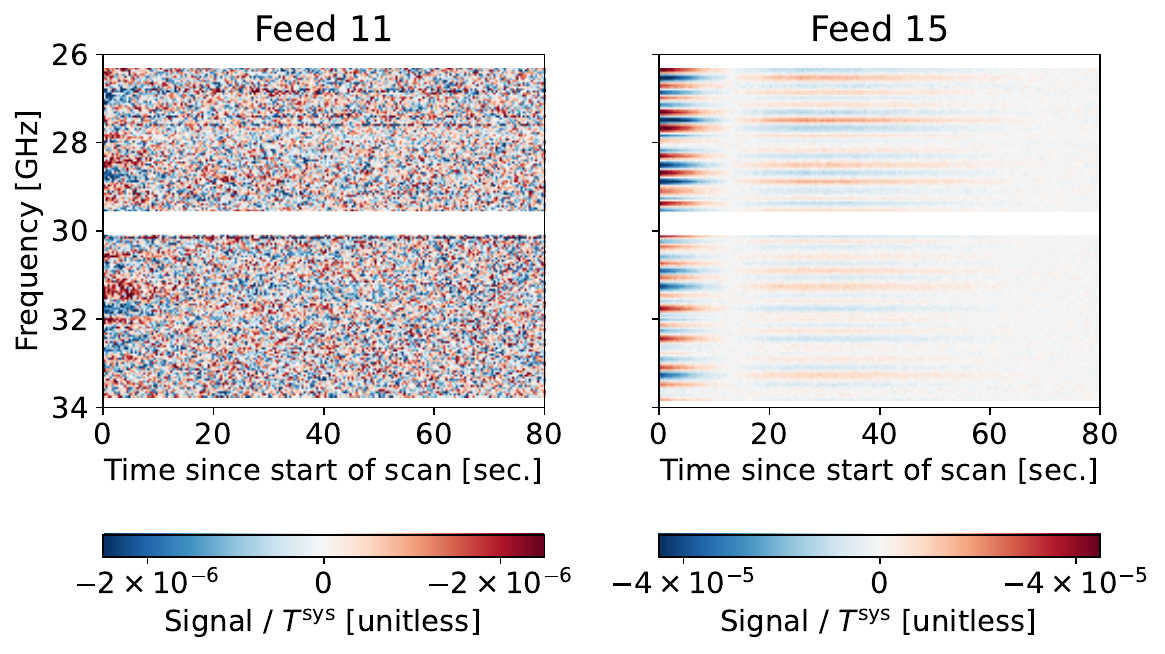}
    \vspace{-4mm}\caption{Same setup as Figure \ref{fig:turnaround-TOD}, but stacked on the beginning of each scan, to emphasize the start-of-scan systematic effect. Notice the difference in the colorbar limits.}
    \label{fig:startofscan-TOD}
\end{figure}

The start-of-scan effect is similar to the turn-around effect in that it is also weak in individual scans, but coherently adds as we add scans. Figure \ref{fig:startofscan-TOD} shows a plot similar to what was presented in the previous section, but that stacks all available Level~2 scans on the beginning of each scan. For Feed 15, we see a very strong wave across frequency, that falls to zero around 17 seconds after the start of the scan, and then switches from negative to positive, or positive to negative power. This is simply an artifact of the low-pass normalization we perform during TOD processing, and the strong wave at the very beginning of each scan is the real start-of-scan feature. This artifact explains why we, in the second and third rows of Figure \ref{fig:mPCA_figures} (where the start-of-scan effect can be seen), observe a similar switching of power as we move from the right edge of the map and toward the center.

Looking at Feed 11 in Figure \ref{fig:startofscan-TOD} a much weaker, but similar, start-of-scan feature can be seen. Generally, all feeds show very similar behavior to either Feed 11 or 15: all Feeds associated with DCM1-2 (Feeds 6, 14, 15, 16, and 17) have very similar behavior, and all remaining Feeds show only a weak start-of-scan effect, as appears in Feed 11. It is unclear why this clear divide exists, and how it relates to DCM1-2.

The exact origin of the start-of-scan effect is unknown, but a standing wave induced by mechanical vibration is also a strong candidate for this systematic effect. The re-pointing that is performed in between scans is currently the only time in the scanning strategy the elevation drive is utilized, as our scans are performed in constant elevation mode.

As with the turn-around effect, some effort was made to model the start-of-scan effect in the time domain. This was fairly successful, and fitting a decaying exponential function to the beginning of each scan appeared to remove more than 90\,\% of the signal induced by this effect. However, the map-PCA proved much more effective than the time-domain efforts, and they are therefore not employed.

\section{Signal injection example maps}\label{app:signal-injection}
\begin{figure}[h!]
  \includegraphics[width=\linewidth]{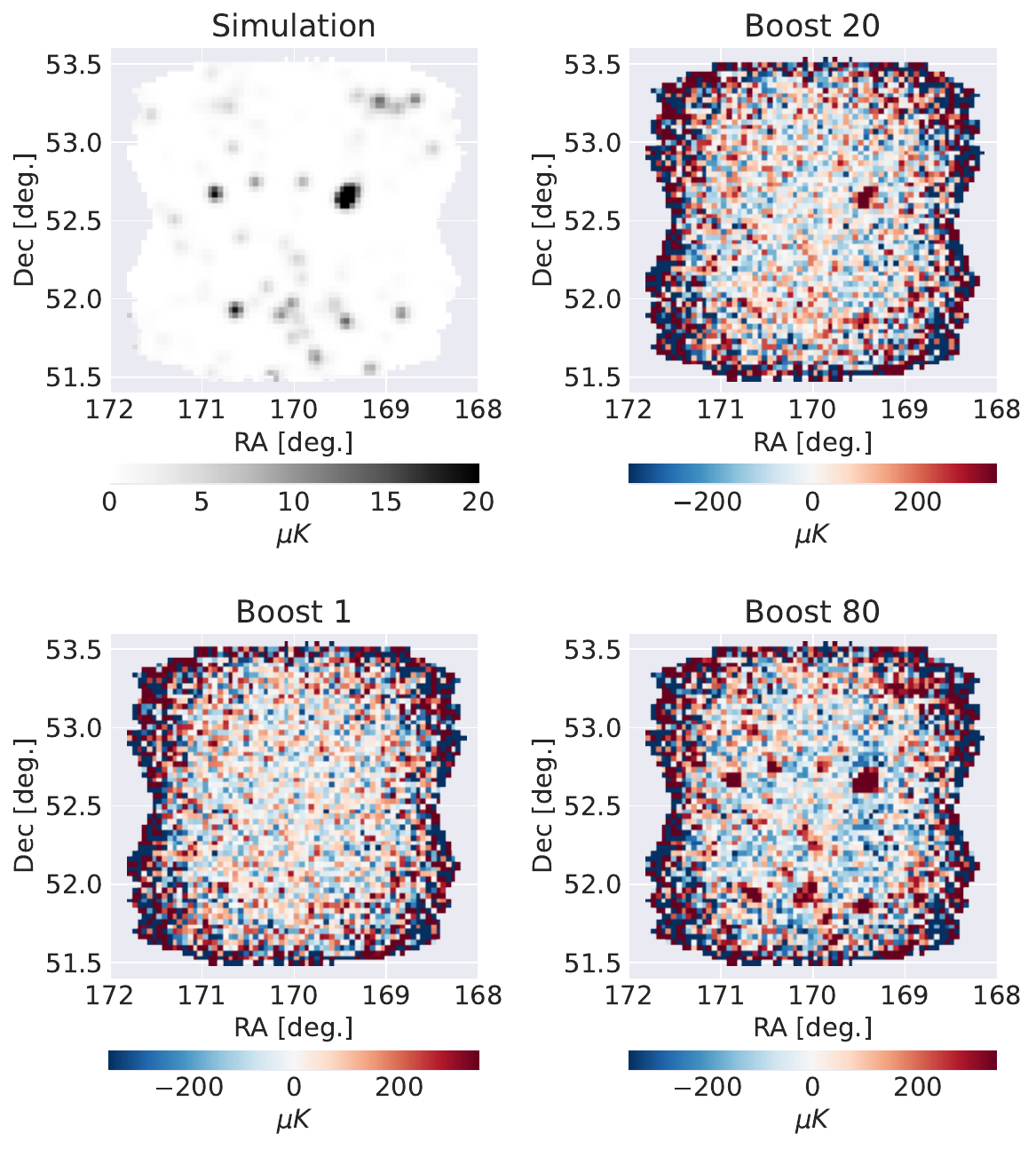}
    \vspace{-4mm}\caption{Illustration of the signal-injection method for transfer function analysis for different signal boost strengths. The top left plot shows the signal-only CO simulation over the relevant patch. The remaining plots show the resulting maps of the simulation injected into the real TOD with different boost strengths and passed through the entire COMAP pipeline. The boost is relative to the fiducial model of \cite{comap-V}, used in the simulations. All four plots show the same frequency slice centered at $\SI{26.953}{GHz}$.}\label{fig:signal_injection_map_example}
\end{figure}

Figure \ref{fig:signal_injection_map_example} illustrates the signal injection pipeline. The first figure shows the CO simulation itself, while the subsequent three panels show the results of injecting this simulation into the real data with different boost strengths. We note that the simulation is injected into the TOD of the Level 1 data, and the maps shown have gone through the entire pipeline.

\section{Uncertainty and frequency maps}
\label{app:maps}

\begin{figure}
    \centering
    \includegraphics[width = \linewidth]{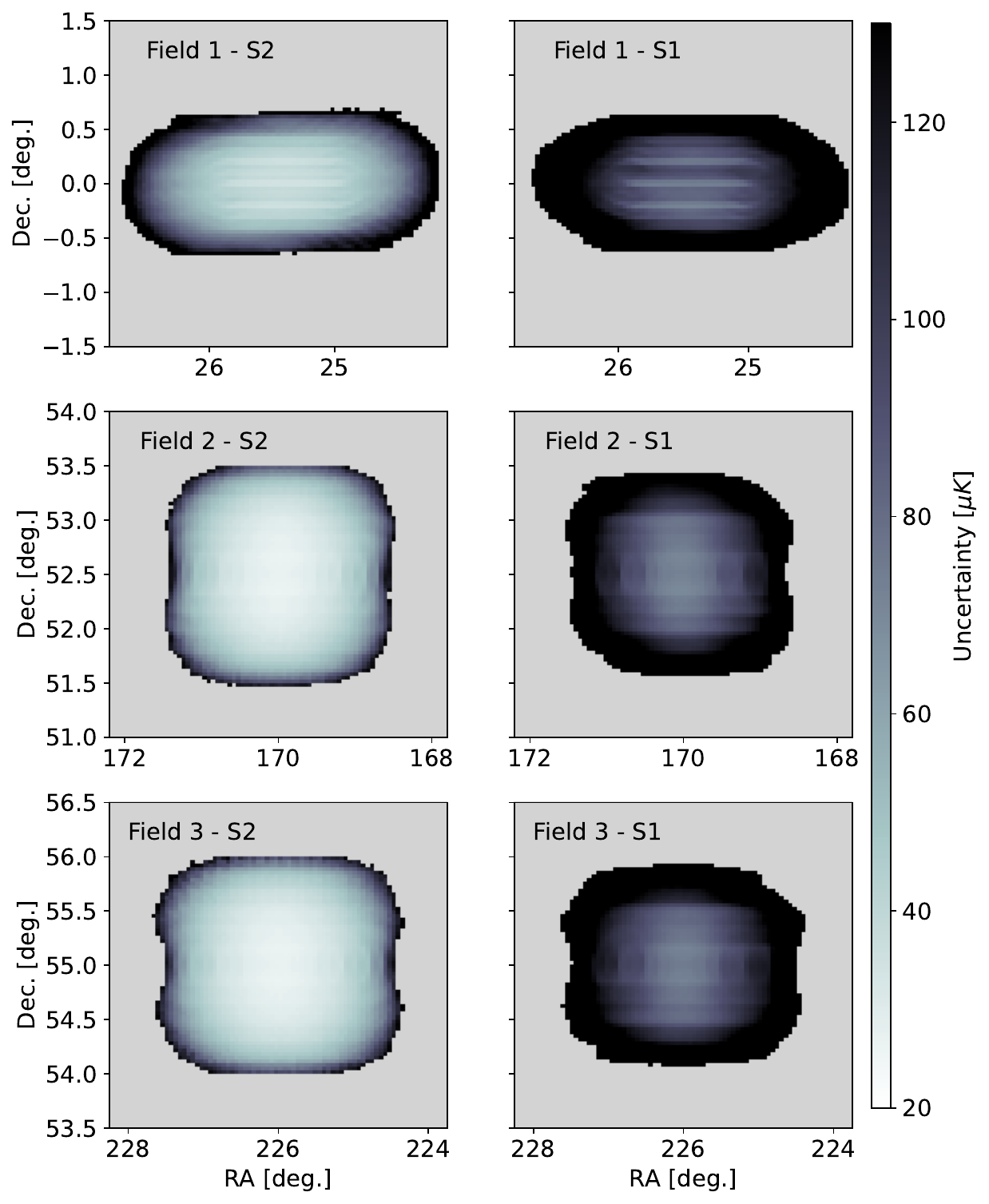}
    \vspace{-4mm}\caption{Uncertainties across the three fields for Season 2 (S2) maps published in this work (left), and the Season 1 (S1) maps published in ES (right).}
    \label{fig:map_rms}
\end{figure}

Figure \ref{fig:map_rms} compares the uncertainties of the Season 2 maps to the Season 1 maps. The values are averages across all frequencies, calculated by inverse-variance co-addition of the uncertainties, as
\begin{equation}
    \sigma^\mathrm{mean} = \sqrt{\frac{1}{\langle 1/\sigma_\nu^2\rangle}}.
\end{equation}
All frequencies have relatively similar uncertainties, with some exceptions close to the Band edges. The center of the maps have a uncertainties of around $\SI{25}{\mu K}$, while the high-sensitivity ${\sim}1.5^\circ \times 1.5^\circ$ regions have an uncertainty of $<\SI{50}{mu K}$ for Fields 2 and 3, with the Field 1 region being slightly smaller.

\begin{figure*}
    \centering
    \includegraphics[width = \linewidth]{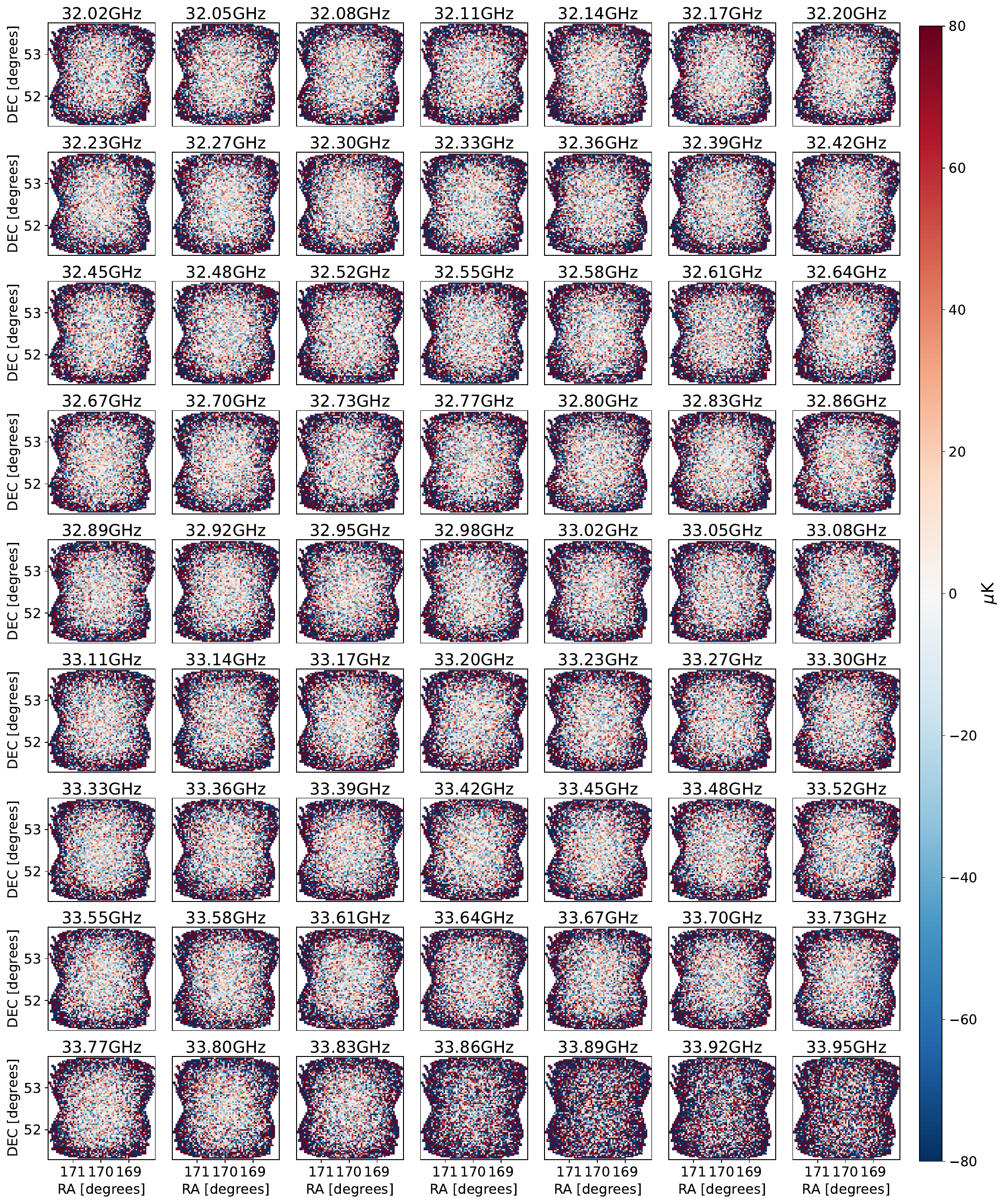}
    \vspace{-4mm}\caption{Final $\SI{31.25}{MHz}$ wide frequency maps for sideband A:USB and Field 2. The titles of each sub-plot indicate the center frequencies of each frequency map. The maps are shown for completeness, and no interesting features beyond the noise can be seen in the maps.}
    \label{fig:maps_USB-A}
\end{figure*}

Figure \ref{fig:maps_USB-A} shows feed-coadded individual frequency maps for Field 2 across $\SI{32}{GHz} - \SI{34}{GHz}$ (1/4th of all channels) for all the data of Season 1 and 2 combined, processed with the Season 2 pipeline. All maps are noise dominated after the map-level PCA filtering. The noise increases toward the highest frequencies because of aliasing cuts on older data (see Sect.~\ref{sec:aliasing}).

\end{document}